\documentclass[twocolumn]{aastex631}

\hypersetup{
    linkcolor=blue,
    citecolor=blue,
    urlcolor=blue
}

\usepackage{amsmath}
\usepackage{comment}
\usepackage{graphicx}
\usepackage{multirow}
\usepackage{natbib}

\newcommand{\harrisonEighteen}{\hyperlink{cite.Harrison_2018}{H18}}
\shorttitle{Spin-Driven Collapsar Jet Breakout}
\shortauthors{ R. Sonawane, K. Kumar \& S. Iyyani}

\begin{document}

\title{From Choked to Successful: The Role of Black Hole Spin in Collapsar Jet Breakout}

\author[0009-0009-2018-9457]{Rushikesh Sonawane}
\affiliation{Centre for High Performance Computing, Indian Institute of Science Education and Research Thiruvananthapuram, 695551, India}
\affiliation{School of Physics, Indian Institute of Science Education and Research Thiruvananthapuram, 695551, India}

\author[0009-0004-0222-788X]{Krishna Kumar}
\affiliation{Department of Applied Mathematics, University of Leeds,
Woodhouse, Leeds LS2 9JT, United Kingdom}
\affiliation{School of Physics, Indian Institute of Science Education and Research Thiruvananthapuram, 695551, India}

% \author[0000-0003-3115-2456]{Ore Gottlieb}
% \affiliation{Center for Computational Astrophysics, Flatiron Institute, 162 5th Avenue, New York, NY 10010, USA}
% \affil{Department of Physics and Columbia Astrophysics Laboratory, Columbia University, Pupin Hall, New York, NY 10027, USA}
% \affil{Department of Physics and Kavli Institute for Astrophysics and Space Research, Massachusetts Institute of Technology, Cambridge, MA 02139, USA}

\author[0000-0003-3220-7543]{Shabnam Iyyani}
\affiliation{School of Physics, Indian Institute of Science Education and Research Thiruvananthapuram, 695551, India}
\affiliation{Centre for High Performance Computing, Indian Institute of Science Education and Research Thiruvananthapuram, 695551, India}

\correspondingauthor{\\
Rushikesh Sonawane: \href{mailto:rushikesh23@iisertvm.ac.in}{rushikesh23@iisertvm.ac.in}\\
Krishna Kumar: \href{mailto:K.Kumar@leeds.ac.uk}{k.kumar@leeds.ac.uk}\\
}



\begin{abstract}

Long gamma-ray bursts (LGRBs) are believed to originate from the core collapse of massive Wolf–Rayet stars, leading to the formation of a spinning black hole that powers a relativistic jet. The prompt gamma-ray emission is produced once the jet successfully propagates through and emerges from the stellar envelope. We perform two-dimensional axisymmetric relativistic hydrodynamic simulations using 
the PLUTO code to model semi-self-consistent, accretion-powered jets launched by Kerr black holes. For the two progenitor models considered, we adopt black hole masses of $4 M_{\odot}$ and $5 M_{\odot}$ corresponding to Wolf–Rayet stars of $10 M_{\odot}$ 
and $25 M_{\odot}$, respectively. 
Unlike previous studies, our simulations employ continuous jet injection, with the injected power linked to the black hole spin and accretion through an empirical relation motivated by GRMHD simulations, enabling a controlled exploration of jet energetics over $10^{48}$--$10^{52}\,\mathrm{erg\,s^{-1}}$.
%Unlike previous studies, the jet power is linked to black hole spin through an empirical relation motivated by GRMHD simulations, allowing controlled exploration of jet energetics across $10^{48}$–$10^{52}\,\mathrm{erg \,s^{-1}}$. 
We investigate jet breakout for Wolf–Rayet progenitors of $10\,M_{\odot}$ and $25\,M_{\odot}$ and identify a critical spin threshold: jets are choked for $a \leq 0.001$ and successfully break out for $a > 0.001$. The breakout time ($t_B$) shows a clear 
correlation with jet luminosity ($L_{\rm jet}$) and spin ($a$), revealing three regimes: Newtonian ($a \lesssim 0.03$), relativistic ($a \gtrsim 0.3$), and an intermediate regime. A distinct dichotomy is observed in the jet head velocity at breakout, with $\beta_h 
\gtrsim 0.8$ for high-spin cases and $\beta_h \sim 0.35$–$0.8$ for lower spins. In the Newtonian regime, simulations yield $t_B \propto L_{\rm jet}^{-0.48}$, steeper than analytical predictions, while the absolute breakout times are systematically longer, indicating modified jet-head scaling. In the relativistic regime, analytical models overestimate breakout times, but calibration improves agreement, with simulations consistently indicating efficient jet propagation in terms of jet breakout.

\end{abstract}

%% Keywords should appear after the \eandnd{abstract} command. 

\section{Introduction}

The propagation of relativistic hydrodynamic jets has been extensively investigated through both analytical and numerical approaches over several decades. Early analytical studies \citep{blandfordrees74, begelman1989, meszaros2001, Matzner_2003, Bromberg_2007, lazzati2005} established the theoretical framework for jet dynamics and their interaction with the surrounding medium, which was later explored in detail through numerical simulations \citep{marti1995, marti1997, aloy2000, macfadyen2001, reynolds2001, zhang2004, mizuta2006, Morsony_2007, wang2008, lazzati2009, mizuta2009, morsony2010, nagakura2011, lazzati2012, lopez-camara2013, ito2015, lopezcamara2016, Gottlieb2018, Gottlieb2020, gottlieb2021structure}. These studies have shown that as a jet propagates through a dense medium, it drives a bow shock into the ambient material, inflating a hot cocoon of shocked plasma that surrounds and collimates the jet, thereby playing a crucial role in determining its dynamics and breakout conditions \citep{Bromberg_11,Harrison_2018,gottlieb2021,Gottlieb_2022_b, Hamidani_2017, Hamidani_2025, Hamidani_2021, Duffell_2015, Duffell_2018,Margalit_2018,Lyutikov_2020}. This physical picture is directly relevant to relativistic outflows associated with long-duration gamma-ray bursts (GRBs), which are widely considered to originate from the core collapse of massive stars \citep{1993ApJ...405..273W,MacFadyen1999,2008ApJ...683L...9C}. Observational evidence linking a subset of long GRBs to broad-lined Type Ic supernovae (e.g., GRB 980425 / SN 1998bw; GRB 030329 / SN 2003dh) has firmly established the GRB–SN connection \citep{1998Natur.395..670G,2003ApJ...591L..17S,2003Natur.423..847H}. In this framework, relativistic jets are launched by a compact central engine—most likely a rapidly rotating black hole \citep{Blandford1977, Popham1999, MacFadyen1999}, or alternatively a highly magnetized, rapidly rotating neutron star \citep{Raynaud2020, Masada2022, White2022}—and must propagate through the dense stellar envelope before breaking out into the circumstellar medium and producing the observed gamma-ray emission \citep{nagakura2012population, Mizuta_and_Ioka_2013, gottlieb_2019, Gottlieb2022c, Gottlieb_2023, urrutia2023three, Urrutia_26}.

Despite significant progress, most studies adopt simplified assumptions about the jet, typically treating it as having a constant luminosity and a fixed opening angle \citep{2003ApJ...586..356Z, 2007ApJ...665..569M}. However, in realistic collapsar systems, the jet power is expected to depend on the mass accretion rate, the magnetic flux of the compact object, and the spin of the central compact object, particularly in black hole–powered systems \citep{Blandford1977, Popham1999, MacFadyen1999}, and the jet angle and orientation evolve over time as well \citep{Gottlieb2022c}. Recent semi-analytical models have begun to incorporate this dependence by explicitly linking jet luminosity to the black hole spin \citep{Lowelle_24}. However, existing studies of jet breakout decouple the jet power from the mass accretion rate, black hole spin, and magnetic flux on the black hole, leaving their role in regulating jet propagation through the stellar envelope unexplored. This is largely due to the multi-scale nature of the problem, in which spin affects the jet near the central engine via general relativistic magnetohydrodynamic (GRMHD) processes, while jet propagation and breakout occur on much larger scales, typically modeled using relativistic hydrodynamics (RHD). A fully self-consistent GRMHD treatment that follows the jet from launch to breakout is computationally expensive, and consequently, most studies adopt simplified engine prescriptions.
In this study, we couple the stellar structure and black hole spin to a continuously injected jet power, with the latter determined by the black hole spin and accretion. This provides a direct connection between the properties of the central engine, the sustained jet energetics, and the resulting breakout conditions of relativistic jets in long-duration GRBs.
To this end, we perform a series of two-dimensional (2D) axisymmetric RHD simulations using the \textsc{PLUTO} code (\citealt{mignone2007pluto}). Rather than explicitly modeling the jet launching process via GRMHD simulations, we prescribe the jet luminosity as a function of both the mass accretion rate and black hole spin, following the semianalytical formalism of \citealt{Lowelle_24}. This approach enables us to incorporate key central engine dependencies in a physically motivated manner while remaining computationally tractable. Using this framework, we systematically explore jet breakout for different spin values in $10\, M_{\odot}$ and $25\, M_{\odot}$ Wolf–Rayet progenitors, and compare our results with analytical expectations to calibrate relations between jet dynamics, breakout time, black hole spin, and the stellar structure.

% \textcolor{orange}{What implications will it have in one-two line?}

This paper is organized as follows: Section \ref{Numerical_Method} describes the numerical setup, including the stellar models and jet launching method. Section \ref{Results} presents the simulation results and their physical implications. Section \ref{Discussion} compares the results with analytical predictions, and Section \ref{Summary} summarizes our findings and discusses their implications.

\section{NUMERICAL METHOD}
\label{Numerical_Method}

\subsection{Stellar Model}

\begin{figure*}[ht!]
\centering

\gridline{
\fig{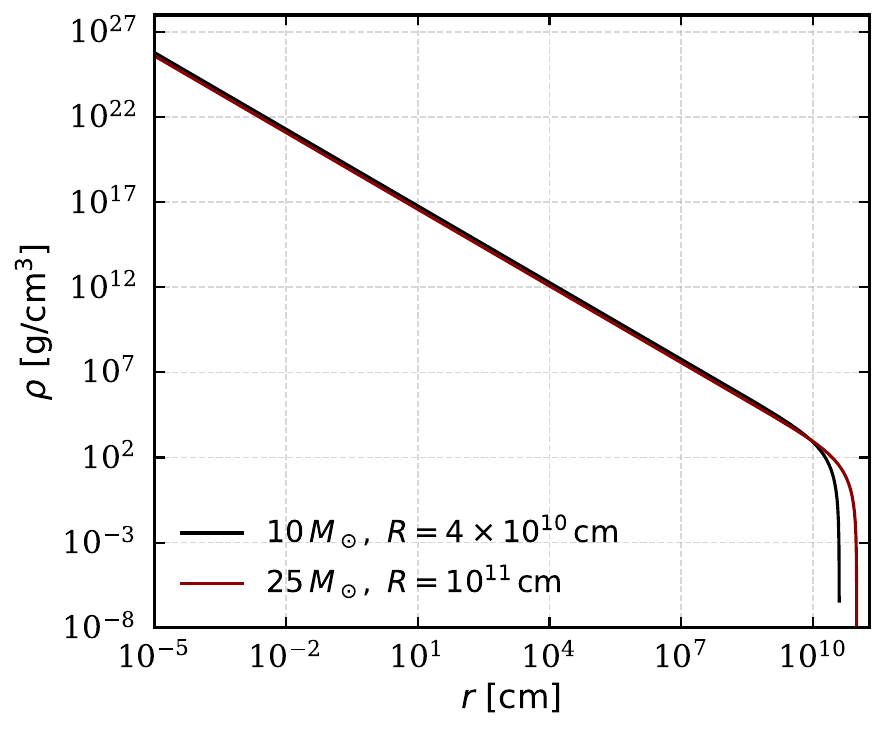}{0.48\textwidth}{(a)}
\fig{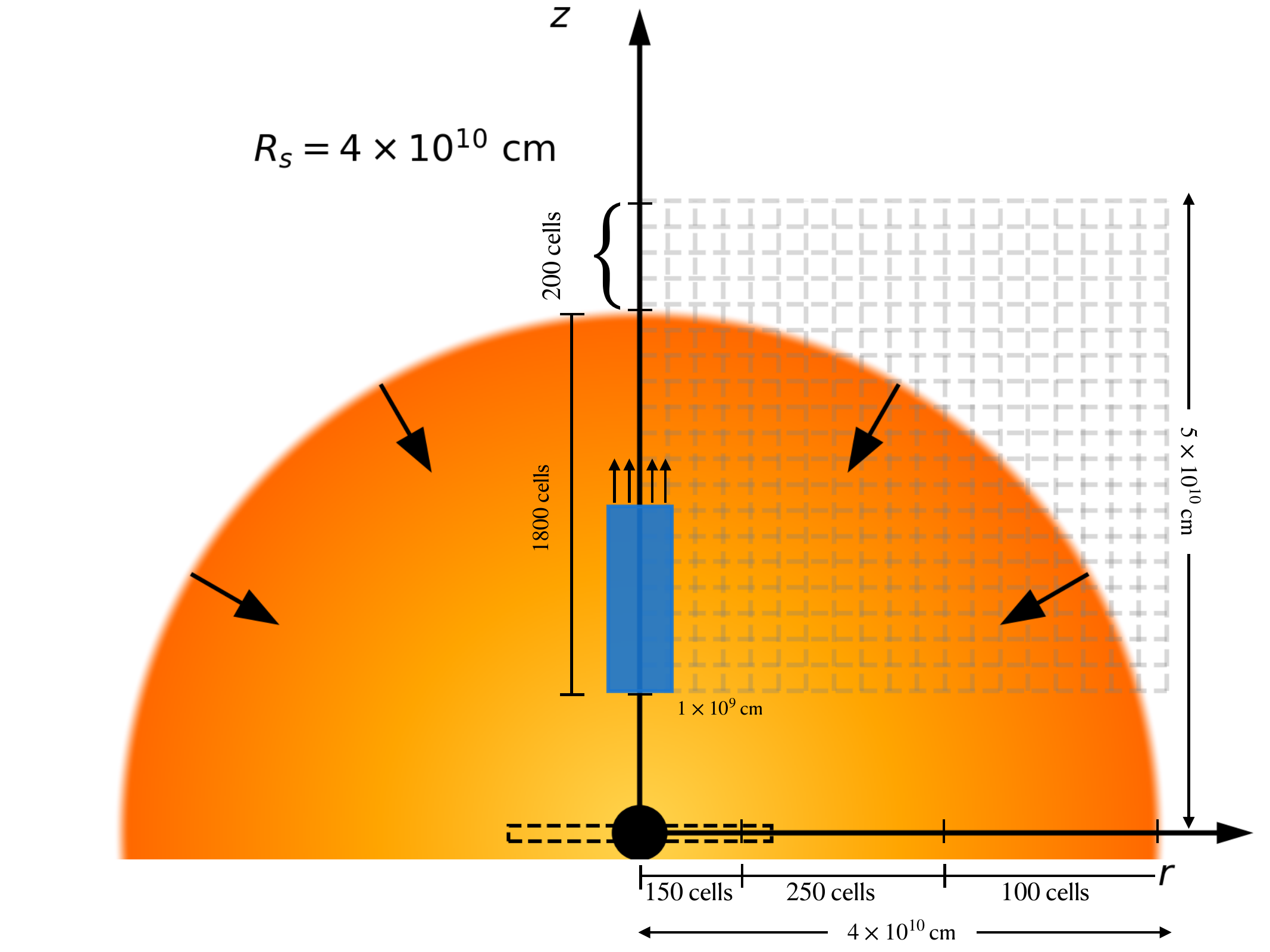}{0.48\textwidth}{(b)}
}

\caption{
(a) Density profiles of Wolf--Rayet stars assuming a power-law slope
$\alpha = 1.5$. The black curve corresponds to a progenitor with stellar mass
$M_{s} = 10\,M_\odot$ and radius $R_{s} = 4\times10^{10}\,\mathrm{cm}$, while the
red curve corresponds to a more massive Wolf--Rayet star with
$M_{s} = 25\,M_\odot$ and radius $R_{s} = 1\times10^{11}\,\mathrm{cm}$.
(b) A schematic of the numerical setup used in our simulations for a $10\,M_\odot$ progenitor star. The star extends up to a radius $R_s$. A relativistic jet is injected along the polar ($z$) axis from an inner boundary located at $r = 1 \times 10^9\,\mathrm{cm}$. The blue rectangular region represents the injected jet, and the upward arrows indicate its propagation. The dashed rectangular box denotes the accretion boundary, which surrounds the black hole acting as the central engine, represented by a filled circle. The larger arrows indicate the radial infall of stellar material. The computational grid is also shown, highlighting the domain size and resolution, with finer zoning near the jet injection region and along the axis.
}

\label{fig:fig1}

\end{figure*}

% We consider a long GRB relativistic jet propagating through the dense envelope of a progenitor Wolf–Rayet star. The stellar envelope is assumed to be unmagnetized, static, and non-rotating. We adopt two progenitor models in this work. The first model corresponds to a $10\,M_{\odot}$ star with a stellar radius $R_{s} = 4 \times 10^{10}$ cm, while the second model corresponds to a $25\,M_{\odot}$ star with a larger radius $R_{s} = 1 \times 10^{11}$ cm. The density profile of the stellar envelope is assumed to be
% \begin{equation}\label{eqn_4}
% \rho(r) = A_{\rho} \, r^{-\alpha} \left(\frac{R_{s} - r}{R_{s}}\right)^3,
% \end{equation}
% where $\alpha$ determines the slope (steepness) of the density profile and $A_{\rho}$ is the normalization constant. For the $10\,M_{\odot}$ progenitor, $A_{\rho} = 1.95 \times 10^{18}$ g cm$^{-1}$, while for the $25\,M_{\odot}$ progenitor, $A_{\rho} = 1.23 \times 10^{18}$ g cm$^{-1}$.
% The pressure inside the star is given by $p = \rho / 10^6$. Outside the star, the ambient medium is modeled with a constant density and pressure given by $\rho_{\rm ISM} = 10^{-14}\,{\rm g\,cm^{-3}}$ and $p_{\rm ISM} = \rho_{\rm ISM}/10^6$, respectively \citep{Bromberg_11,Harrison_2018}.
% Figure~\ref{fig:fig1} shows (a) the . density profile of the Wolf–Rayet star for $\alpha = 1.5$, and (b) a schematic diagram illustrating the stellar collapse and the jet injected at a radius of $10^9$~cm.

We consider a collapsar relativistic jet propagating through the dense envelope of a progenitor star. The stellar envelope is set to be unmagnetized, static, and non-rotating. We adopt two progenitor models in this work. The first model corresponds to a $10\,M_{\odot}$ star with a stellar radius $R_{s} = 4 \times 10^{10}$ cm, while the second model corresponds to a $25\,M_{\odot}$ star with a larger radius $R_{s} = 1 \times 10^{11}$ cm. The density profile of the stellar envelope follows a free-fall profile during collapse \citep{halevi2023density}, given by
\begin{equation}\label{eqn_4}
\rho(r) = A_{\rho} r^{-1.5} \left(\frac{R_{s} - r}{R_{s}}\right)^3,
\end{equation}
where $A_{\rho}$ is the normalization constant. For the $10\,M_{\odot}$ progenitor, $A_{\rho} = 1.95 \times 10^{18}\,\mathrm{g\,cm^{-1.5}}$, while for the $25\,M_{\odot}$ progenitor, $A_{\rho} = 1.23 \times 10^{18}\,\mathrm{g\,cm^{-1.5}}$. The pressure inside the star is given by $p = \rho c^2 / 10^6$. Outside the star, the ambient medium is modeled with a constant density and pressure given by $\rho_{\rm ISM} = 10^{-14}\,\mathrm{g\,cm^{-3}}$ and $p_{\rm ISM} = \rho_{\rm ISM}c^2/10^6$, respectively. Figure~\ref{fig:fig1} shows (a) the density profile and (b) illustrates a schematic diagram of the stellar collapse and the jet launched at a radius of $10^9$~cm.
% where $A_{\rho}$ is the normalization constant. For the $10,M_{\odot}$ progenitor, $A_{\rho} = 1.95 \times 10^{18}$ g cm$^{-1.5}$, while for the $25,M_{\odot}$ progenitor, $A_{\rho} = 1.23 \times 10^{18}$ g cm$^{-1.5}$. The pressure inside the star is given by $p = \rho / 10^6$. Outside the star, the ambient medium is modeled with a constant density and pressure given by $\rho_{\rm ISM} = 10^{-14},{\rm g,cm^{-3}}$ and $p_{\rm ISM} = \rho_{\rm ISM}/10^6$, respectively \citep{Bromberg_11,Harrison_2018}. Figure~\ref{fig:fig1} shows (a) the density profile of the Wolf–Rayet star, and (b) a schematic diagram illustrating the stellar collapse and the jet injected at a radius of $10^9$~cm.

% \begin{figure}[H]
% \centering
% \includegraphics[width=1\textwidth]{Schematic+density.pdf}
% \caption{
% (a) Density profiles of Wolf--Rayet stars assuming a power-law slope
% $\alpha = 1.5$. The black curve corresponds to a progenitor with stellar mass
% $M_{s} = 10\,M_\odot$ and radius $R_{s} = 4\times10^{10}\,\mathrm{cm}$, while the
% red curve corresponds to a more massive Wolf--Rayet star with
% $M_{s} = 25\,M_\odot$ and radius $R_{s} = 1\times10^{11}\,\mathrm{cm}$.
% (b) Schematic illustration of the collapse of a Wolf--Rayet star and the
% launching of a long gamma-ray burst (GRB) jet. The computational grid used in
% the $10\,M_\odot$ simulation is also shown.
% }

% \label{fig1}
% \end{figure}

\subsection{Jet Injection Mechanism}

We inject a relativistically hot cylindrical jet with an initial Lorentz factor $\Gamma_{j,0}$ through a nozzle of radius $r_{\rm noz}$ located at a height $z_{\rm base}$ along the $z$-axis. Following \citet[hereafter H18]{Harrison_2018}, the jet, owing to its high internal energy, undergoes rapid acceleration and lateral expansion, eventually forming a cold conical outflow with an opening angle $\theta_0 \approx 1/(f\Gamma_{j,0})$, where $f \approx 1.4$. We adopt $r_{\rm noz} = 10^8$ cm and $z_{\rm base} = 10^9$ cm, as in \harrisonEighteen, to ensure adequate numerical resolution near the injection region. The jet is initialized with $\Gamma_{j,0} = 10$, corresponding to $\theta_0 \approx 0.071$ rad, and a specific enthalpy $h_{j,0} = 100$. The jet density and pressure at injection are given by
\begin{equation}
\rho_j = \frac{L_{jet}}{\pi r_{\rm noz}^2 \Gamma_{j,0}^2 c^3 h_{j,0}}, 
\quad
p_j = \frac{(h_{j,0} - 1)\rho_j}{4}.
\label{eqn_7}
\end{equation}

We adopt an accretion-powered jet luminosity mechanism rather than injecting a constant jet luminosity into the star–jet system. The jet is continuously injected through the nozzle, with its luminosity and thermodynamic properties evolving in time according to the accretion rate.

A successful launch of a relativistic Blandford–Znajek (BZ) jet requires the formation of an accretion disk around the central black hole, which depends on the angular momentum of the infalling material \citep{MacFadyen1999}. Independently, we assume that the accretion timescale of the stellar envelope is governed by the free-fall time. Accordingly, a shell accreted at time $t$ originates from an initial radius
\begin{equation}
r_0(t) = (2GM_{BH}t^2)^{1/3},
\end{equation}
where $M_{BH}$ is the black hole mass.

The corresponding free-fall mass supply rate to the disk is given by
\begin{equation}
\begin{aligned}\label{eq:Mdot}
\dot{M}_{d}(t, \alpha) &= \frac{8\pi \rho_0 r_0(t)}{3R^3 t} \biggl[ \\
&\quad r(t)^{(5 - \alpha)} + 3 R r(t)^{(4 - \alpha)} \\
&\quad - 3 R^2 r(t)^{(3 - \alpha)} + R^3 r(t)^{(2 - \alpha)} \biggr],
\end{aligned}
\end{equation}
where $r_0(t)$ is the initial radius of the infalling shell \citep{Gottlieb_2022}.

In the case of a Kerr black hole in the Magnetically Arrested Disk (MAD) state, where the magnetic flux saturates, the jet power is determined by the mass accretion rate and a spin-dependent efficiency \citep{tchekhovskoy2015magnetic, Lowelle_24}. The jet luminosity is then given by
\begin{equation}
L_{jet} = \eta(a)\,\dot{M}_d c^2,
\end{equation}
where the accretion rate is set by the progenitor’s density profile according to Eq.~\eqref{eq:Mdot}, and $\eta(a)$ is the electromagnetic efficiency, which is determined by the dimensionless black hole spin $a$ \citep{Lowelle_24},
\begin{equation}
\eta(a) = 1.063a^4 + 0.395a^2.
\end{equation}

\begin{figure}[hb!]
\centering
\includegraphics[width=1.05\linewidth]{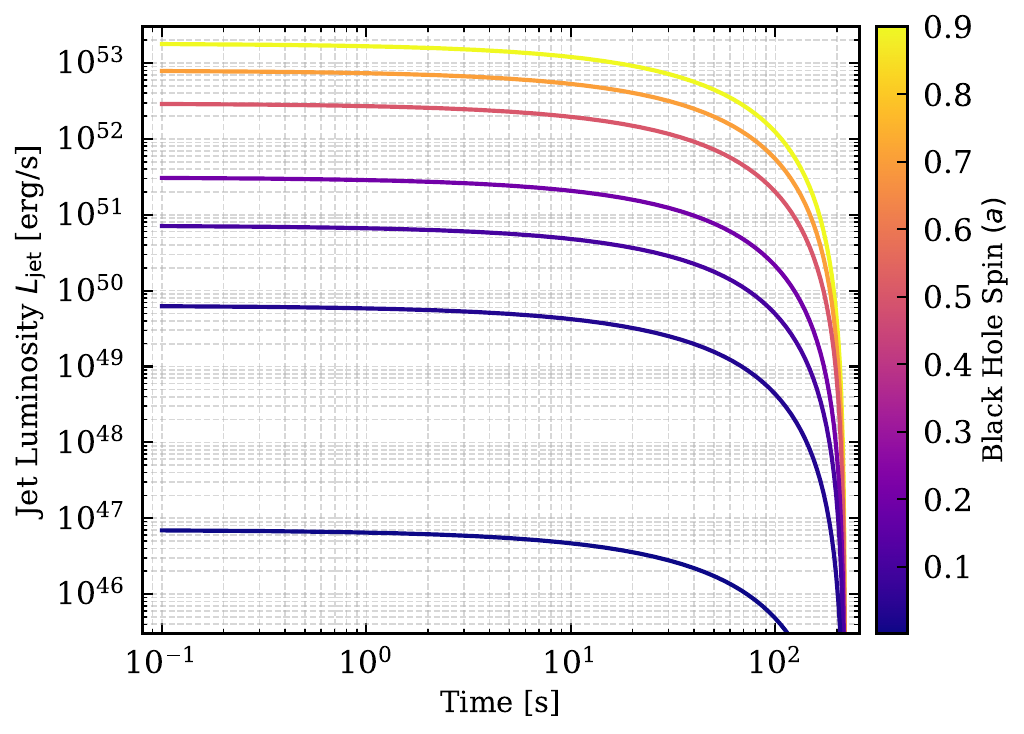}
\caption{
Evolution of jet luminosity with time for different black hole spin values for a $10\,M_{\odot}$ progenitor star.
}
\label{fig:fig2}
\end{figure}

Figure~\ref{fig:fig2} shows the semi-analytic evolution of the accretion-powered jet luminosity for different spin values. The sharp decline in luminosity at $t \sim 250$ s corresponds to the free-fall timescale of the stellar envelope for a $10\,M_{\odot}$ progenitor star. In the MAD regime, efficiencies exceeding $100\%$ indicate extraction of rotational energy from the black hole \citep{tchekhovskoy2011black,mckinney2012general}. A qualitatively similar behavior is obtained for a $25\,M_{\odot}$ progenitor.

While the jet power is derived assuming magnetically driven outflows that become hydrodynamically dominated at the injection point, the jet propagation is modeled self-consistently.
For each simulation, we assume a constant black hole spin parameter $a$. While the spin can evolve due to ongoing accretion and jet launching—particularly in the high-spin regime (e.g., \citealt{Jacquemin-Ide_2024}), modeling this evolution is beyond the scope of the present study.
\\
In summary, our simulation setup involves the continuous injection of jet power, extracted from the black hole spin and accretion, until either the jet successfully breaks out of the star or the stellar free-fall time is reached. Within this framework, we investigate the breakout times corresponding to different jet powers, with particular focus on the influence of black hole spin.

\subsection{Simulation Grid}
% The simulations presented in this work have been carried out using the PLUTO code (version 4.0) on a static grid z-r in two-dimensional cylindrical coordinates. On the r-axis, one uniform patch with 150 cells to 1.0 $\times 10^9$ cm, a logarithmic patch with 250 cells to 2 $\times 10^{10}$ cm, and an outer logarithmic patch with 100 cells to 4.01 $\times 10^{10}$ cm have been used. On the z-axis, we employ one uniform patch with 1800 cells from $z_{beg}$ to $ 4 \times$ $10^{10}$ cm and an outer logarithmic patch with 200 cells to 5 $\times 10^{10}$ cm. In total, our 2D grid includes 500 $\times$ 2000 cells. Figure~\ref{fig1}(b) shows the grid structure used in the simulations, indicating the ranges along both axes and the number of cells in each region.

% {\bf include the info for 25 solar mass star}

The simulations presented in this work were carried out using the \textsc{PLUTO} code (version 4.0) on a static two-dimensional $z$--$r$ grid in cylindrical 
coordinates. Along the $r$-axis, the grid consists of three patches: a uniform patch with 150 cells extending to $1.0 \times 10^{9}\,\mathrm{cm}$, followed by a logarithmic patch with 250 cells extending to $2.0 \times 10^{10}\,\mathrm{cm}$, and an outer logarithmic patch with 100 cells extending 
to $4.01 \times 10^{10}\,\mathrm{cm}$. Along the $z$-axis, we employ a uniform patch with 1800 cells extending from $z_{\mathrm{base}}$ to $4.0 \times 10^{10}\,\mathrm{cm}$, followed by an outer logarithmic patch with 200 cells extending to $5.0 
\times 10^{10}\,\mathrm{cm}$. In total, the two-dimensional computational domain comprises $500 \times 2000$ grid cells. Figure~\ref{fig:fig1}(b) illustrates the grid structure used in the simulations, indicating the spatial extent and number of cells along both axes.

Similarly, for the $25\,M_{\odot}$ progenitor star, the grid along the $r$-axis consists of a uniform patch with 250 cells extending to $1.0 \times 10^{9}\,\mathrm{cm}$, followed by a logarithmic 
patch with 350 cells extending to $3.0 \times 10^{10}\,\mathrm{cm}$, and an outer logarithmic patch with 200 cells extending to $1.04 \times 
10^{11}\,\mathrm{cm}$. Along the $z$-axis, we employ a uniform patch with 2100 cells extending from $z_{\mathrm{base}}$ to $1.0 \times 
10^{11}\,\mathrm{cm}$, followed by an outer logarithmic patch with 300 cells extending to $1.3 \times 10^{11}\,\mathrm{cm}$. The resulting two-dimensional grid for this case consists of $800 \times 2400$ cells.
\section{RESULTS}
\label{Results}
% \subsection{Conditions for Jet Breakout and Choking}
% The breakout condition is determined by the emergence of relativistic gas from the star, typically observed sometime after the forward shock breakout. Figure \ref{fig:rho_spins} shows the logarithmic density map of an accretion-powered jet breaking out of the WR stellar envelope for high spin BH [right most], moderate spin [ middle panel], and low spin BH [left panel]. It has been observed that hyper-accreting high-spin BH produces long GRBs with higher luminosity, while most long GRBs of an observed characteristic luminosity are produced by slowly spinning BH. We observed that the breakout time for high-spin BH-powered jets is relatively much smaller than for low-spin BH-powered jets. It’s noteworthy that even after propagating for 26 seconds inside the stellar core, the jet remains choked at a spin of $a$ = 0.01 (Figure \ref{fig: Chocked Jets}(a)). If the simulations are allowed to evolve for another ten seconds, there might be a possibility of the jets breaking out. However, whether this would result in classical GRBs is a question of interest for further studies. Our numerical simulations have effectively constrained the feasible BH spins capable of powering relativistic jets.

\subsection{Conditions for Jet Breakout and Choking}

\begin{figure*}
\centering

% -------- Top row --------
\gridline{
\fig{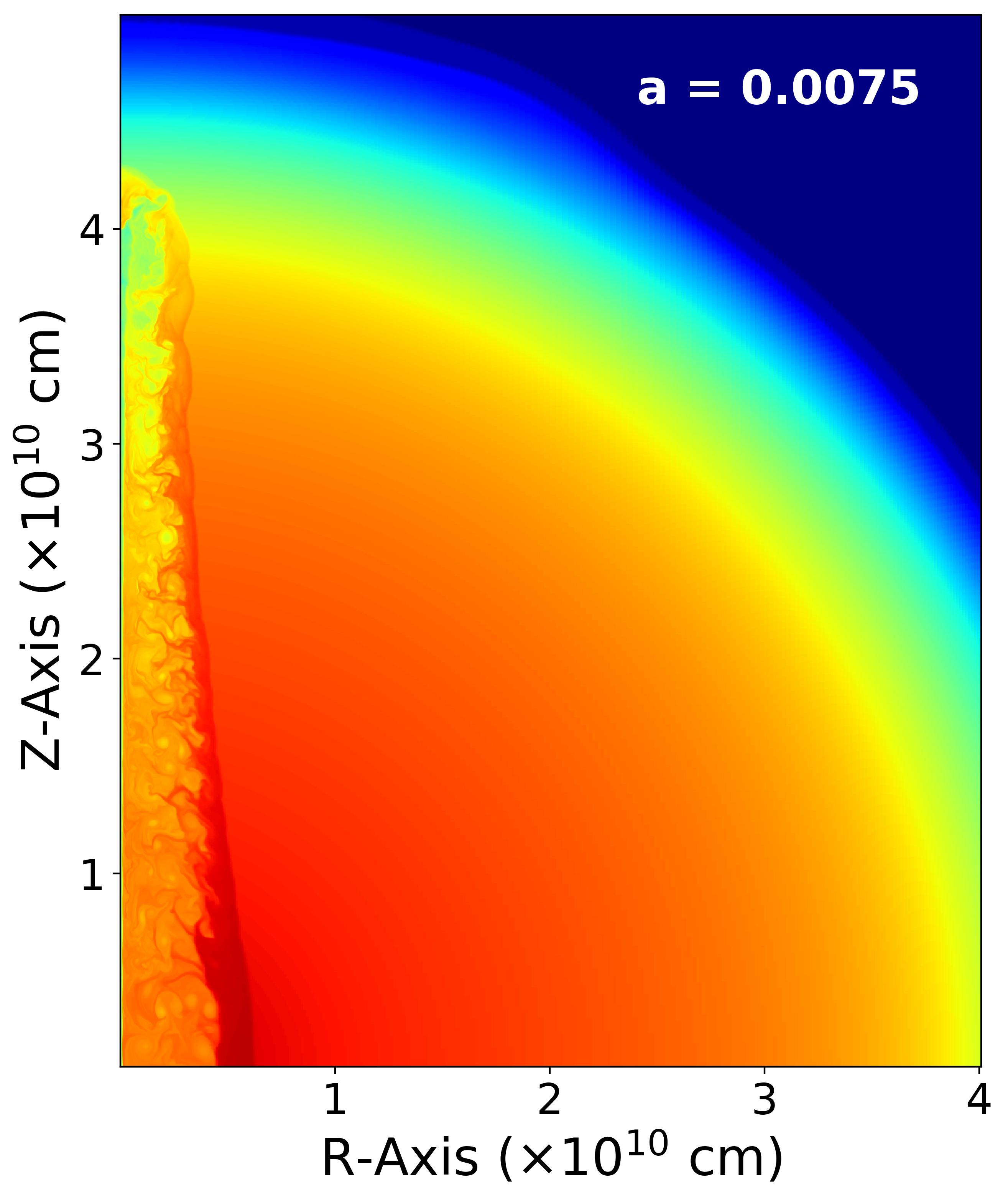}{0.308\textwidth}{(a)}
\fig{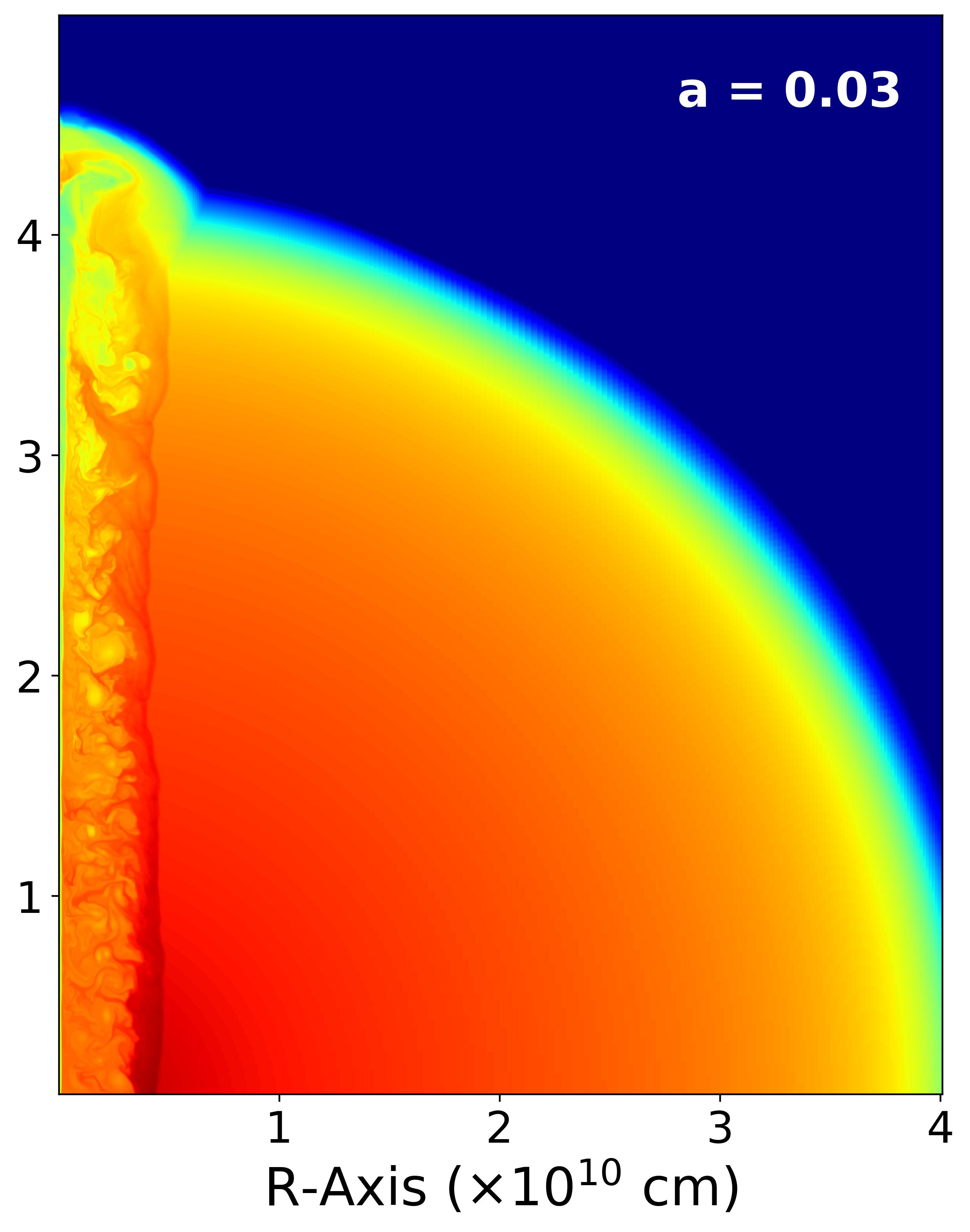}{0.29\textwidth}{(b)}
\fig{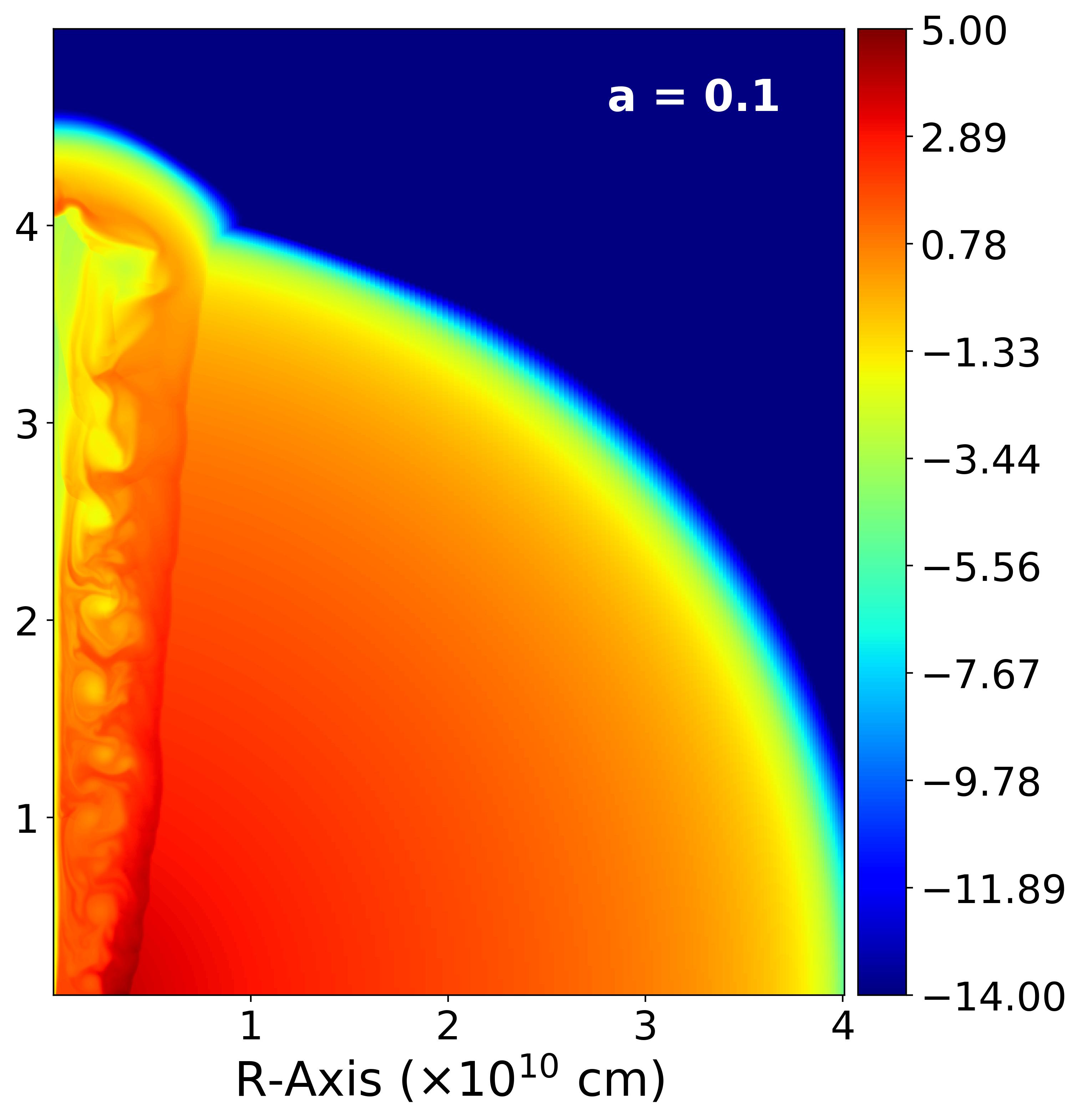}{0.36\textwidth}{(c)}
}

\vspace{0.3cm}

% -------- Bottom row --------
\gridline{
\fig{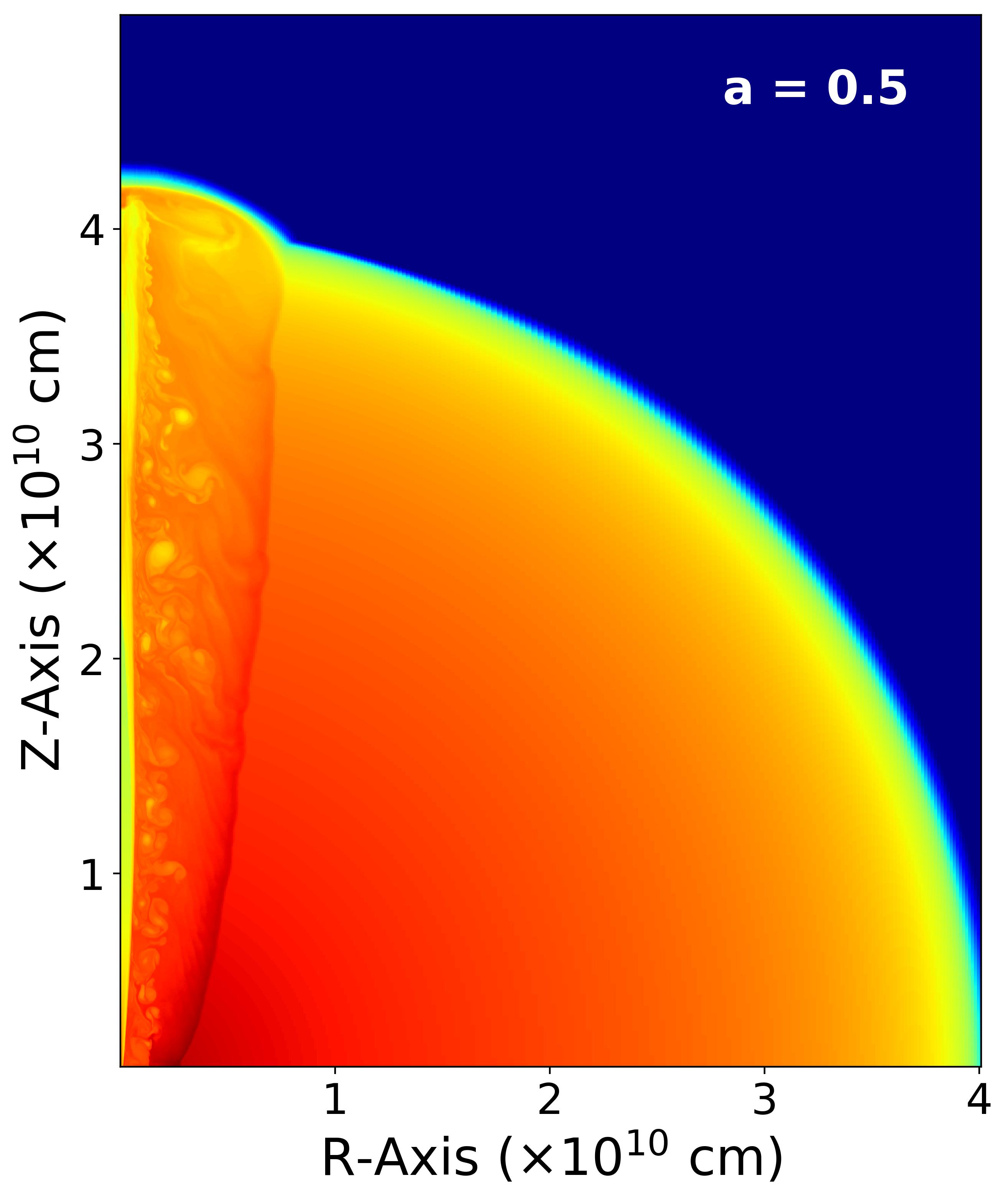}{0.308\textwidth}{(d)}
\fig{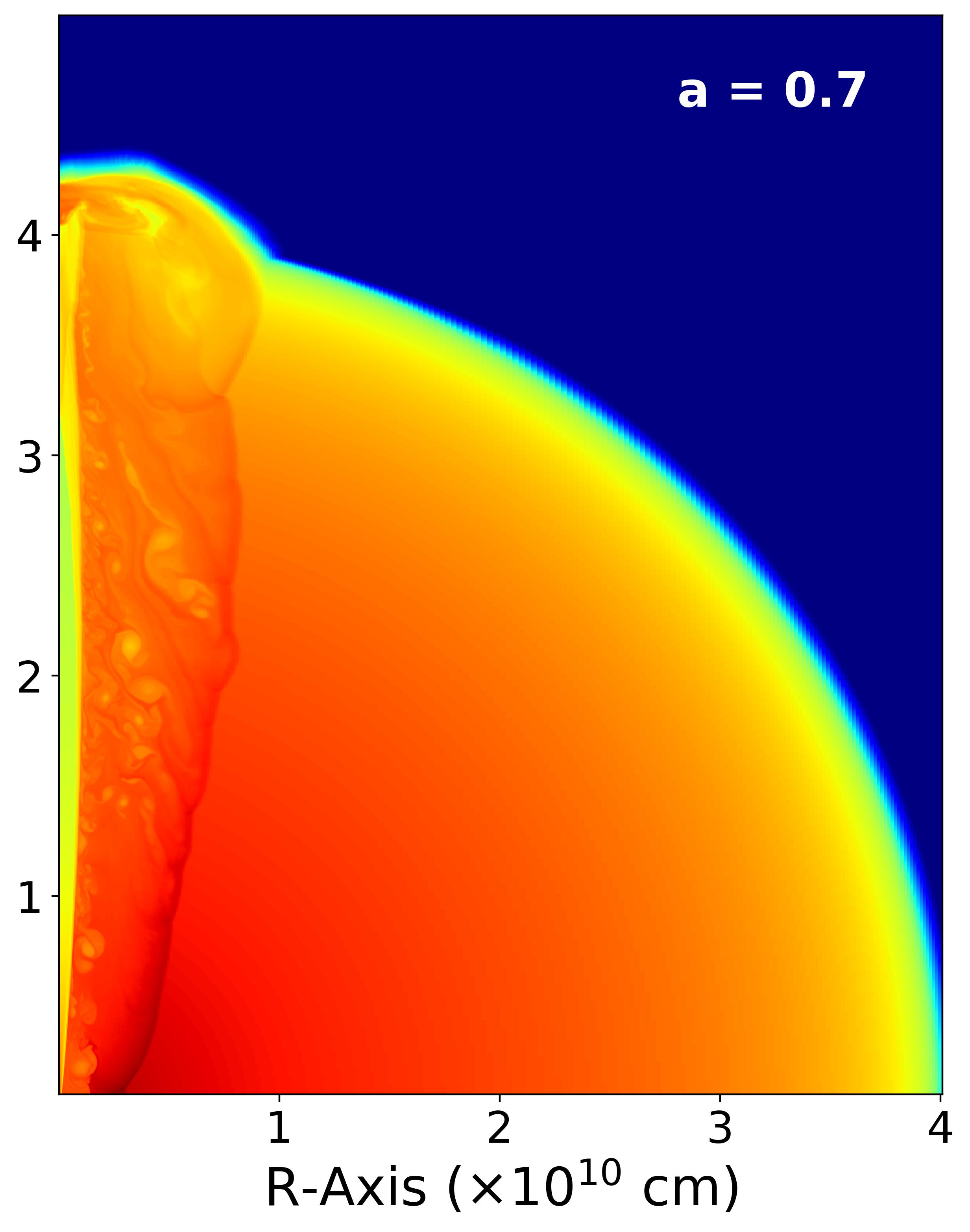}{0.29\textwidth}{(e)}
\fig{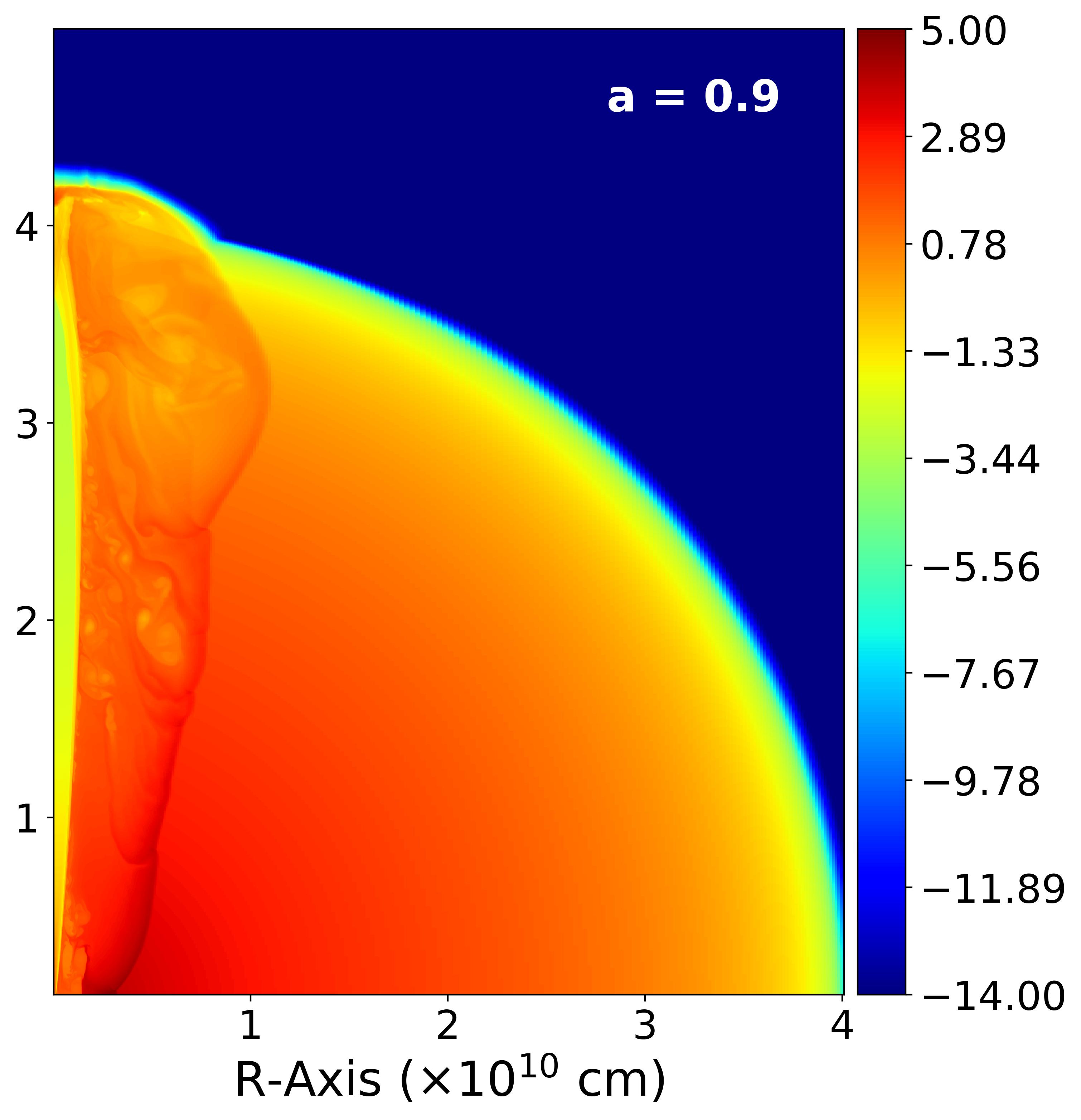}{0.36\textwidth}{(f)}
}

\caption{
Logarithmic density maps (in $\mathrm{g\,cm}^{-3}$) of collapsar jets upon breakout from a $10\,M_{\odot}$ Wolf--Rayet progenitor star for different black hole spins. 
\textbf{Top row (a–c):} Newtonian regime with spins $a = 0.0075$, $0.03$, and $0.1$. 
\textbf{Bottom row (d–f):} Relativistic regime with spins $a = 0.5$, $0.7$, and $0.9$. 
The colorbar corresponds to $\log_{10}(\rho)$.
}

\label{fig:rho_spins_combined_10}

\end{figure*}

The breakout of a relativistic jet from a collapsar progenitor is determined by its successful emergence from the stellar envelope (\citealp{Bromberg_11}, \harrisonEighteen). During its propagation, the jet drives a forward shock into the ambient stellar material and a reverse shock into the jet, forming a shocked jet head that advances by balancing the jet ram pressure against the confining pressure of the surrounding medium, while inflating a hot cocoon that influences collimation. The nature of this propagation can be characterized by the dimensionless parameter $\tilde{L}$\footnote{The dimensionless parameter $\tilde{L} = L_{jet} / (\Sigma_j \rho_a c^3)$, where $L_{jet}$ is the jet luminosity, $\Sigma_j$ is the jet cross-sectional area of the jet head, $\rho_a$ is the ambient density, and $c$ is the speed of light}, it quantifies the ratio of jet energy density to the ambient rest-mass energy density at the jet head. For $\tilde{L} \ll 1$, the jet head propagates sub-relativistically (\textit{Newtonian regime}), whereas for $\tilde{L} \gg 1$, it moves relativistically (\textit{relativistic regime}). A successful breakout does not require continuous engine activity, provided that the information about engine shutdown carried by the jet tail does not reach the jet head before breakout; otherwise, the loss of pressure support can cause the jet to stall.

Figure~\ref{fig:rho_spins_combined_10} depicts the logarithmic density maps of accretion-powered jets propagating through the stellar envelope at breakout. Here, we focus on the $10\,M_{\odot}$ progenitor, while the corresponding results for the $25\,M_{\odot}$ case are presented in Appendix~\ref{sec:appendix_25Msun}. The jet morphology spans a range of black hole spins covering both Newtonian and relativistic regimes, with low-spin cases and high-spin cases. The simulations show that jet propagation and collimation strongly depend on the black hole spin, which regulates the jet luminosity $L_{jet}$. As the spin increases, the jets become more powerful, propagate more efficiently through the stellar envelope, and carve a more pronounced low-density polar channel.
% Figures \ref{fig:rho_spins_combined_10} and \ref{fig:rho_spins_combined_25} depicts the logarithmic density maps of accretion-powered jets propagating through the stellar envelope upon breakout for two progenitor masses of $10\,M_{\odot}$ and $25\,M_{\odot}$, respectively. 
% Each figure illustrates the jet morphology for a range of black hole spins spanning both the Newtonian and relativistic regimes. The top rows correspond to low-spin cases ($a = 0.0075$, $0.03$, and $0.1$), while the bottom rows show higher-spin cases ($a = 0.5$, $0.7$, and $0.9$). The simulations 
demonstrate that jet propagation and collimation strongly depend on the spin of the central black hole, which regulates the jet launching efficiency and the resulting jet luminosity, $L_{jet}$. As the spin increases, the jets become more powerful and propagate more efficiently through the stellar envelope, producing a more pronounced low-density channel along the polar direction.

We model a continuously powered jet, with energy injection from the central engine maintained throughout the simulation. Each model is evolved until either the jet breaks out of the stellar surface or the stellar free-fall timescale, $t_{\rm ff}$, is reached. In the majority of cases considered in this study, the jet successfully breaks out well before $t_{\rm ff}$. For models that do not achieve breakout within this time, we examine the position and propagation speed of the jet head at $t_{\rm ff}$. In these cases, the jet head remains deeply embedded within the stellar envelope, typically having traversed less than $\sim 25\%$ of the stellar radius, while propagating at an average dimensionless speed of only $\beta_h \sim 0.007$. Such jets show no indication of imminent breakout and are therefore classified as choked jets. To characterize the jet propagation efficiency, we compare the breakout time, $t_{\rm B}$, with the stellar free-fall timescale and present the ratio $t_{\rm B}/t_{\rm ff}$ for the successful breakout models.

\begin{figure*}
\centering

\gridline{
\fig{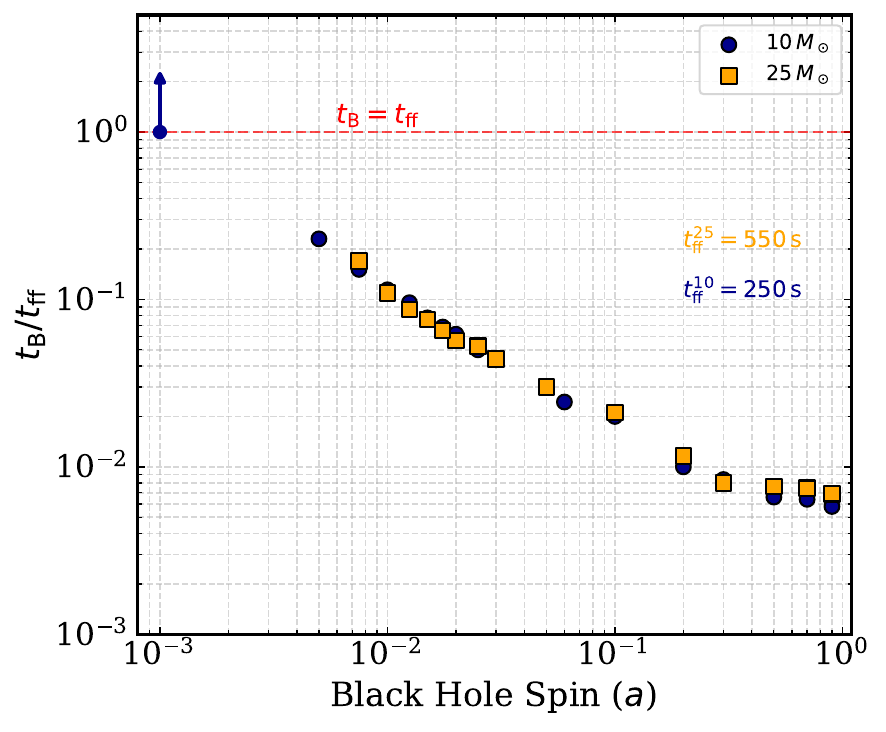}{0.48\textwidth}{(a)}
\fig{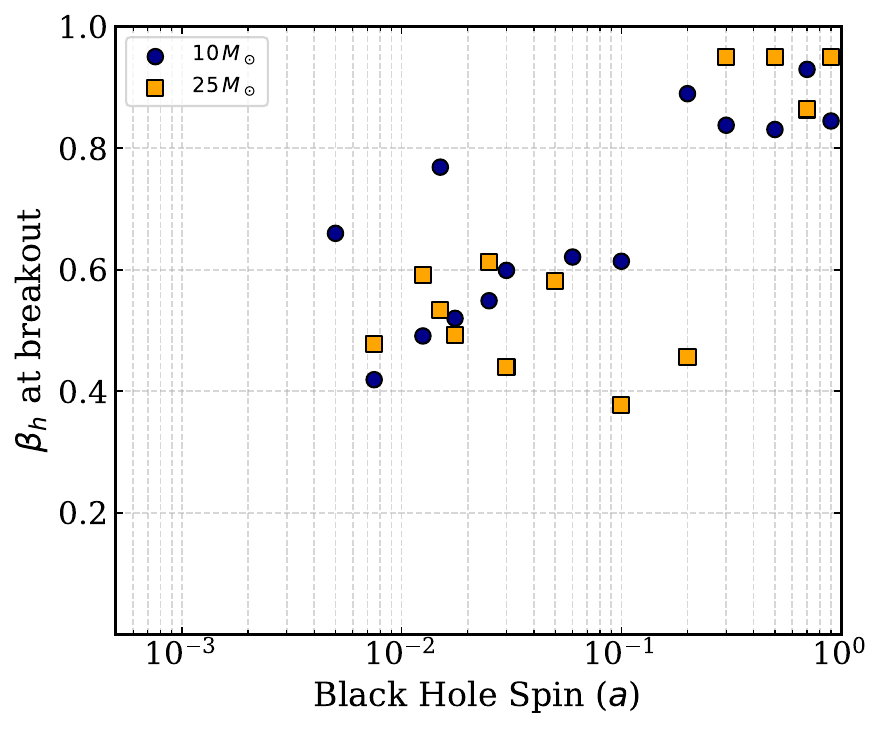}{0.48\textwidth}{(b)}
}

\caption{
\textbf{(a)} Variation of the breakout time normalized by the free-fall time \((t_{\rm B}/t_{\rm ff})\) as a function of black hole spin \(a\). The red dashed line corresponds to \(t_{\rm B} = t_{\rm ff}\). The upward red arrow at \(a = 0.001\) indicates a lower limit on \(t_{\rm breakout}/t_{\rm ff}\) at that spin. 
\textbf{(b)} Jet head velocity ($\beta_h$) as a function of spin at the time of breakout. In both panels, blue circles represent the $10 M_{\odot}$ progenitor, while orange squares denote the $25 M_{\odot}$ progenitor.
}

\label{fig:Chocked_Jet_spin}

\end{figure*}

Figure \ref{fig:Chocked_Jet_spin}(a) shows the variation of the jet breakout timescales, normalized by the stellar free-fall time 
($t_{\rm ff} \sim 250\,\mathrm{s}$ for $10\,M_{\odot}$ and $t_{\rm ff}\sim 550\,\mathrm{s}$ for $25\,M_{\odot}$), as a function of the black hole spin parameter $a$. The breakout times obtained for the different black hole spins for $10\,M_{\odot}$ and $25\,M_{\odot}$ stars are reported in Table \ref{tab:combined}. Over the range of black-hole spins explored in this study ($0.001 \le a \le 0.9$), we note the transition from efficient jet breakout to jet choking. Jets launched by systems with $a \gtrsim 0.01$ successfully emerge from the progenitor star, with breakout times typically below $\sim 10\%$ of the stellar free-fall timescale, $t_{\rm ff}$. As 
the spin decreases towards $a \sim 0.01$, the breakout time increases and the jet approaches the threshold between successful and unsuccessful propagation. For yet lower spins, the reduced jet luminosity leads to inefficient 
propagation through the stellar envelope, ultimately resulting in choked jets. Below, we discuss representative examples of the near-threshold breakout and choked-jet regimes. 

% For the $10\,M_{\odot}$ progenitor, we focus on the jet propagation corresponding to a low spin case ($a=0.01$). As shown in Figure \ref{fig:ChokedJets}(a), the jet remains confined within the star after $\sim 26\,\mathrm{s}$ ($\sim 10\%$ of the free-fall time), with the jet head yet to reach the stellar surface and the relativistic outflow 
% still embedded within the core. Upon extending the simulation by an additional $\sim 10\,\mathrm{s}$, the jet eventually breaks out; however, such a 
% delayed breakout is likely to produce a weaker transient than classical long-duration GRBs, potentially appearing as a low-luminosity GRB (\citealt{Irwin_2016, Irwin_2025}), shock breakout signal, or cocoon-dominated emission (\citealt{Nakar_2017}). This is further supported by the jet head velocity at breakout, $\beta_h$. For successful breakout cases, $\beta_h \gtrsim 0.1$, indicating mildly relativistic to relativistic propagation, whereas for spins $a \lesssim 0.03$, the jet head becomes sub-relativistic (Figure \ref{fig:Chocked_Jet_spin}(b)).

\begin{figure*}
\centering
\includegraphics[width=0.9\textwidth]{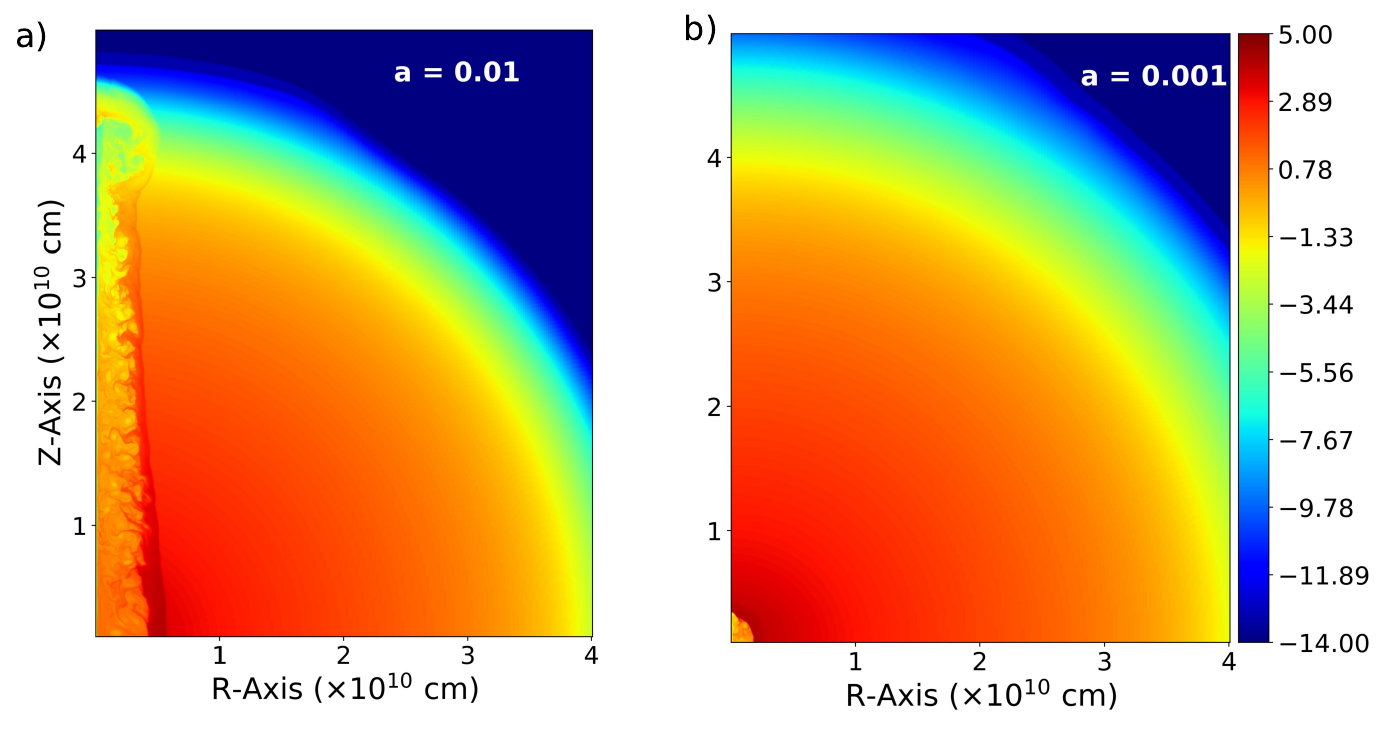}

\caption{For $10\, M_{\odot}$ star: \textbf{Panel (a):} Logarithmic density map for a black-hole spin of $a = 0.01$ at the time = 28.6 s ($\sim 11\%$ of the $t_{\rm ff}$) is shown.
\textbf{Panel (b):} Logarithmic density map for $a = 0.001$ at time = 50 s ($\sim 20\%$ of the $t_{\rm ff}$) is shown, clearly indicating a stalled jet within the stellar core.}

\label{fig:ChokedJets}
\end{figure*}

For the $10\,M_{\odot}$ progenitor, the case with spin $a=0.01$ represents a near-threshold breakout scenario. As shown in Figure~\ref{fig:ChokedJets}(a), at $\sim 28.6\,\mathrm{s}$ (corresponding to $\sim11\%$ of the stellar free-fall time), the jet head remains within the stellar surface, 
with the relativistic outflow still confined to the inner regions of the progenitor and only modestly perturbing the overlying envelope. The jet breaks out of the star $\sim 10\,\mathrm{s}$ later. Owing to its relatively long breakout time compared to the more rapidly emerging jets in our sample, such an event may produce a weaker 
transient than a classical long-duration GRB.  

For an even lower spin, $a=0.001$, the jet luminosity is insufficient to sustain efficient propagation through the stellar envelope. Figure~\ref{fig:ChokedJets}(b) shows the jet structure 
at $\sim 20\%$ of the stellar free-fall time, by which stage the jet head has advanced only a small distance from the injection region. The propagation remains extremely slow throughout the evolution, and even after evolving the 
model up to $t_{\rm ff}$, the jet head remains deeply embedded within the stellar core. Unlike the $a=0.01$ case, there is no indication of imminent breakout, demonstrating that such low-spin systems produce genuinely choked jets.

% \begin{figure*}
% \centering

% \begin{subfigure}[t]{0.48\textwidth}
%     \includegraphics[width=\linewidth]{Time_vs_Spin.pdf}
%     \caption{}
% \end{subfigure}
% \hfill
% \begin{subfigure}[t]{0.48\textwidth}
%     \includegraphics[width=\linewidth]{beta_vs_spin.pdf}
%     \caption{}
% \end{subfigure}

% \caption{\textbf{Panel (a):} Variation of the breakout time normalized by the free-fall time \((t_{\rm B}/t_{\rm ff})\) as a function of black hole spin \(a\). The red dashed line corresponds to \(t_{\rm B} = t_{\rm ff}\). The upward red arrow at \(a = 0.001\) indicates a lower limit on \(t_{\rm breakout}/t_{\rm ff}\) at that spin. \textbf{Panel (b):} Jet head velocity ($\beta_h$) as a function of spin at the time of breakout. The red dashed line marks $\beta_h = 0.1$. In both panels, blue circles represent the $10 M_{\odot}$ progenitor, while orange squares denote the $25 M_{\odot}$ progenitor.}

% \label{fig:Chocked_Jet_spin}

% \end{figure*}

To further characterize the successful breakout cases, we examine the jet-head velocity at the time of breakout (Figure~\ref{fig:Chocked_Jet_spin}(b)). For both the $10\,M_{\odot}$ and $25\,M_{\odot}$ progenitors, a clear dichotomy emerges between the Newtonian and relativistic jet-propagation regimes. Jets in the relativistic regime ($\tilde{L}>1$), typically associated with rapidly spinning black holes ($a\gtrsim0.3$), break out with highly relativistic jet-head velocities, $\beta_h\gtrsim0.8$. In contrast, jets in the Newtonian regime ($\tilde{L}<1$) emerge with substantially lower breakout velocities, $\beta_h\sim0.35$--$0.8$. The observed dichotomy in breakout velocities is consistent with the breakout-time behavior discussed above, with relativistic jets penetrating the stellar envelope more efficiently, while Newtonian jets require longer propagation times before reaching the stellar surface.

\subsection{Breakout Time Correlations with Jet Luminosity and Black Hole Spin}

\label{breakout_corr}
As discussed in the previous subsection, higher BH spins result in shorter jet breakout times. To quantify the trends, we perform a systematic analysis of jet breakout conditions obtained for the different black hole spins in the case of both $10\,M_{\odot}$ and $25\,M_{\odot}$ progenitors. 
%As expected, higher spins result in shorter breakout times (Figure~\ref{fig:combined_breakout}(b)). 
We characterize this dependence by fitting segmented power-law relations between the breakout time, jet luminosity, and black hole spin, with the goodness of fit assessed using the coefficient of determination ($R^2$).

\begin{figure*}[ht]
\centering

\gridline{
\fig{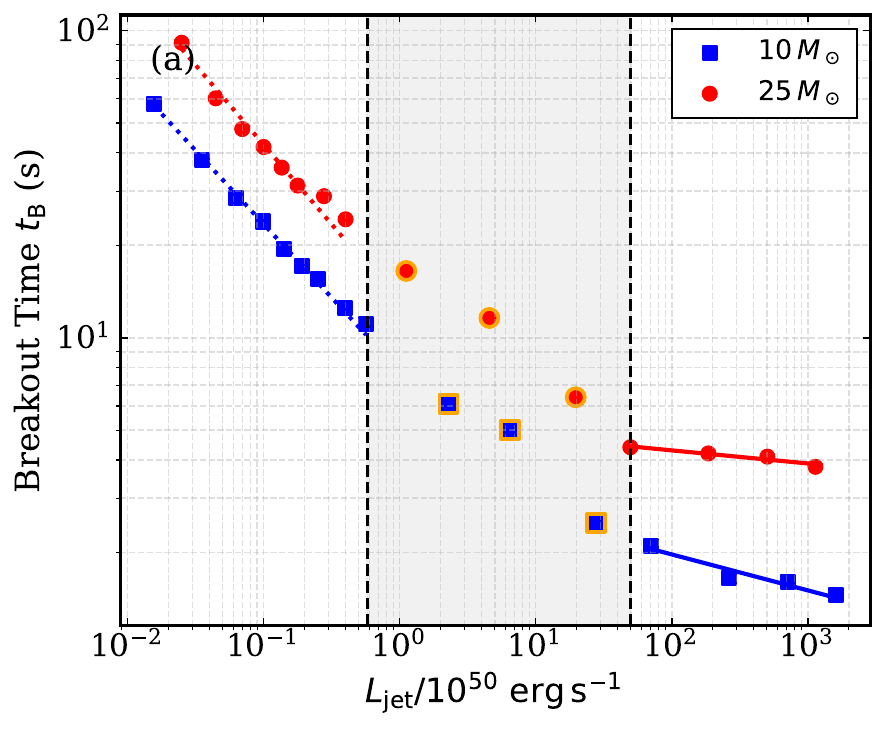}{0.45\textwidth}{(a)}
\fig{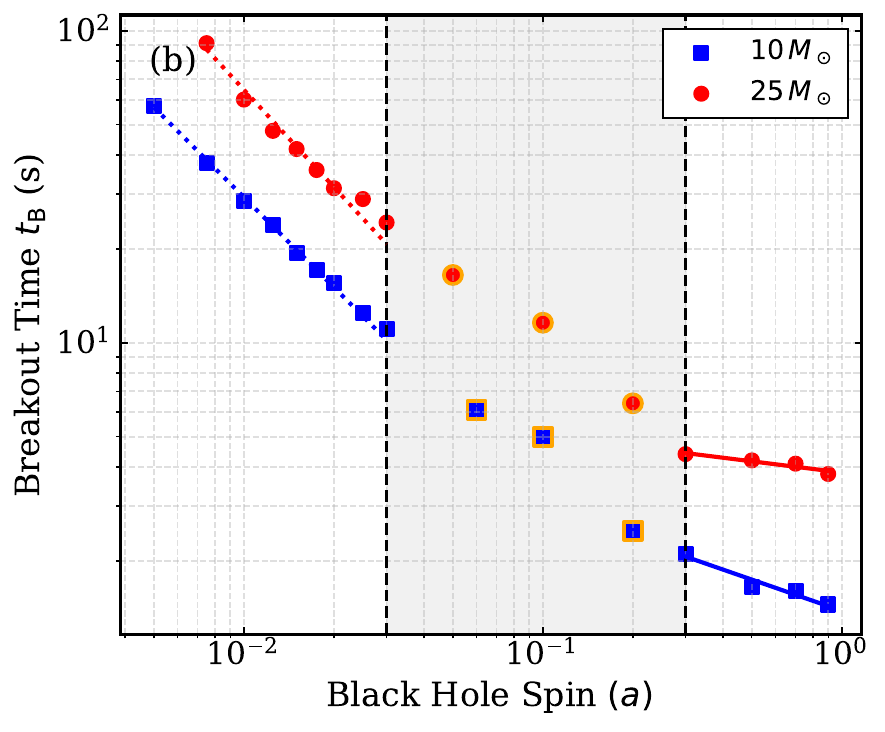}{0.45\textwidth}{(b)}
}

\caption{
\textbf{(a)} Jet breakout time as a function of jet luminosity.
\textbf{(b)} Jet breakout time as a function of black hole spin.
In both panels, segmented power-law fits are shown for jets propagating
through $10\,M_{\odot}$ and $25\,M_{\odot}$ progenitor stars.
Solid lines represent fits in the relativistic regime, while dotted lines
correspond to the Newtonian regime. Square markers denote the
$10\,M_{\odot}$ star, and circular markers denote the
$25\,M_{\odot}$ star.
}

\label{fig:breakout_spin}

\end{figure*}

Figure~\ref{fig:breakout_spin}(b) shows the relationship between
black hole spin and breakout time, while Figure~\ref{fig:breakout_spin}(a)
shows the dependence on jet luminosity. The observable trend reveals three main regimes: Newtonian ($a\le 0.03$; $L_{jet} <6 \times 10^{49} \, \rm erg/s$; ${\tilde L}_a < 1$), relativistic ($a>0.2$, $L_{jet} > 3 \times 10^{52} \, \rm erg/s$; ${\tilde L}_a > 1$), while there is an intermediate transition regime corresponding to BH spins $0.2\ge a > 0.03$. The Newtonian and Relativistic regimes were analysed using segmented power-law fits for both progenitor masses. 

For the $10\,M_\odot$ progenitor, we obtain
\begin{align}
t_B &= 7.79 \times L_{\mathrm{jet}}^{-0.48}
\quad \text{(low-luminosity regime)},
\label{eq:10Msun_low} \\
t_B &= 3.87 \times L_{\mathrm{jet}}^{-0.12}
\quad \text{(high-luminosity regime)},
\label{eq:ll10}
\end{align}
% For the $10\,M_\odot$ progenitor, the best-fit relations are
% \begin{align}
% t_B &= 7.79\times L_{\mathrm{j}}^{-0.48}
% \quad \text{(low-luminosity regime)}
% \label{eq:eq9}\\
% t_B &= 3.87 \times L_{\mathrm{j}}^{-0.14}
% \quad \text{(high-luminosity regime)}
% \label{eq:ll10}
% \end{align}

with coefficients of determination $R^2 = 0.998$ and $0.960$, respectively, indicating good fit to the simulation results.

For the $25\,M_\odot$ progenitor, we obtain
\begin{align}
t_B &= 8.42 \times L_{\mathrm{jet}}^{-0.52}
\quad \text{(low-luminosity regime)},
\label{eq:25Msun_low} \\
t_B &= 4.11 \times L_{\mathrm{jet}}^{-0.04}
\quad \text{(high-luminosity regime)},
\label{eq:25Msun_high}
\end{align}
% For the $25\,M_\odot$ progenitor, we obtain
% \begin{align}
% t_B &= 8.42 \times L_{\mathrm{j}}^{-0.53}
% \quad \text{(low-luminosity regime)}
% \label{eq:25Msun_low} \\
% t_B &= 4.11 \times L_{\mathrm{j}}^{-0.12 }
% \quad \text{(high-luminosity regime)}
% \label{eq:25Msun_high}
% \end{align}

with the corresponding $R^2$ values of $0.973$ and $0.741$. 

% The increased scatter in the high-luminosity regime for the $25\,M_{\odot}$ progenitor likely reflects the enhanced influence of stellar structure and jet–envelope interactions in more massive stars.
Overall, these results demonstrate that a segmented power-law model robustly captures the breakout time across different progenitor masses, while highlighting systematic, mass-dependent deviations at high jet luminosities.

\subsection{Analytical Relation: Calibration in the Newtonian Regime}
\label{Newtonian_calibration}
%To further examine the physical consistency of our simulations, we use the dimensionless parameter $\tilde{L}$ introduced by \citet{Matzner_2003,Harrison_2018}. This parameter represents the ratio of the jet energy density to the ambient rest-mass energy density at the jet head and plays a key role in determining the jet propagation dynamics. 
To assess the physical consistency of our simulations, we employ the dimensionless parameter $\tilde{L}$, which governs the jet propagation dynamics (\citealp{Matzner_2003}, \harrisonEighteen). In the collimated regime, $\tilde{L}$ can be written as (\citealp{Bromberg_11}, \harrisonEighteen):

\begin{equation}
\tilde{L} = \left( \frac{L_{jet}}{\theta_0^4 t^2 \rho_a(z_h) c^5} \right)^{\frac{2}{5}}  
\left( \frac{16 \Omega}{3\pi} \right)^{\frac{2}{5}},
\label{eq:collimated}
\end{equation}

where the factor $\Omega$ depends on the stellar density profile and is given by

\begin{equation}
\Omega =
\begin{cases} 
\frac{81}{(5 - \alpha)^3(3 - \alpha)} & \text{if } \tilde{L} \gg 1, \\[5pt] 
\frac{3(3 + \alpha)^2}{5(3 - \alpha)(7 - \alpha)} & \text{if } \tilde{L} \ll 1.
\end{cases}
\label{Omega}
\end{equation}

The jet head velocity is determined by balancing the ram pressure across the forward and reverse shocks \citep{begelman1989overpressured,Matzner_2003}:
\begin{equation}
\rho_j h_j \Gamma_j^2 \Gamma_h^2 (\beta_j - \beta_h)^2 + P_j =
\rho_a h_a \Gamma_h^2 \beta_h^2 + P_a.
\end{equation}
In the limit of a cold ambient medium and a strong reverse shock ($P_a, P_j \ll$ ram pressure), this reduces to
\begin{equation}
\beta_h = \frac{\beta_j}{1 + \tilde{L}^{-1/2}},
\label{eq:3}
\end{equation}
where
\begin{equation}
\tilde{L} \equiv \frac{\rho_j h_j \Gamma_j^2}{\rho_a}.
\label{eq:4}
\end{equation}

The jet-head velocity $\beta_h$ is related to $\tilde{L}$ through
\begin{equation}
\beta_h = \frac{1}{1 + \tilde{L}^{-1/2}} \approx \tilde{L}^{1/2} \qquad (\tilde{L} \ll 1).
\label{eq 21}
\end{equation}

Using this expression, we can estimate the theoretical relation between the jet breakout time ($t_B$) and the jet luminosity ($L_j$). The propagation of the jet head inside the stellar envelope satisfies

\[
R = c \int_0^{t_B} \beta_h(t)\, dt,
\]

where $R$ is the stellar radius. Substituting $\beta_h \approx \tilde{L}^{1/2}$ and the analytical expression for $\tilde{L}$ (Equation~\ref{eq:collimated}), and performing the integration, we obtain the relation

\[
t_B \propto L_{jet}^{-1/3}.
\]

This represents the theoretical prediction for the dependence of the breakout time on the jet luminosity in the Newtonian regime. However, our simulation results presented in Section \ref{breakout_corr} show a steeper dependence. In particular, we find $t_B \propto L_{\mathrm{jet}}^{-0.48 }$ for the $10\,M_\odot$ progenitor and $t_B \propto 
L_{\mathrm{jet}}^{-0.53 }$ for the $25\,M_\odot$ progenitor. This discrepancy suggests that the simple analytical approximation may not fully capture the jet propagation dynamics in our simulations.

To quantify the agreement between the analytical estimate ($\tilde{L}_a$) and the simulation-derived value ($\tilde{L}_s$), we introduce a calibration factor defined as
\begin{equation}
N_s = \sqrt{\frac{\tilde{L}_s}{\tilde{L}_a}}.
\label{eq:Ns}
\end{equation}

The quantity $\tilde{L}_s$ is estimated using Equation~\ref{eq 21}, where the jet head velocity ($\beta_h$) is measured in the region where it attains a steady value. These $\beta_h$ values are then used to compute $\tilde{L}_s$. The analytical estimate $\tilde{L}_a$ is calculated independently using Equation~\ref{eq:collimated}. For each spin configuration, $N_s$ is evaluated using the corresponding values of $\tilde{L}_s$ and $\tilde{L}_a$.

\begin{figure}[t]
\centering
\includegraphics[width=\columnwidth]{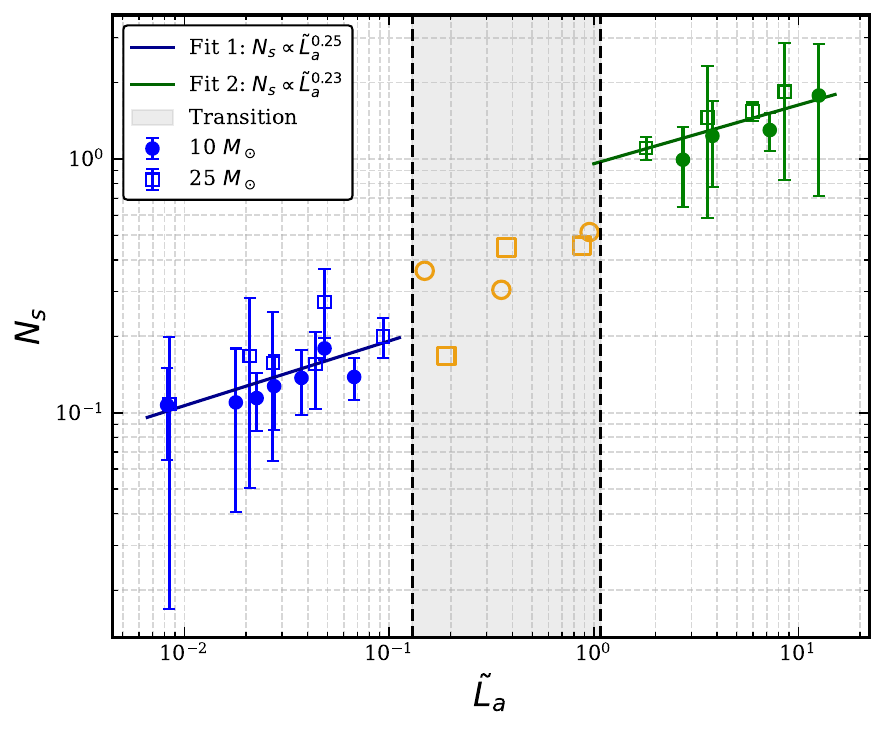}
\caption{
$N_s$ as a function of $\tilde{L}_a$ for 10 and 25~$M_\odot$ models. 
Filled circles (squares) denote 10~$M_\odot$ (25~$M_\odot$) data points. 
Blue and green colors correspond to the \textit{Newtonian} and \textit{Relativistic} regimes, respectively, 
while open markers indicate points in the transition region.
}
\label{fig:calibration}
\end{figure}

The dependence of $N_s$ on $\tilde{L}_a$ is shown in Figure~\ref{fig:calibration}, and the corresponding values of $\tilde{L}_s$ and $\tilde{L}_a$ are listed in Table~\ref{tab:combined}. The data exhibit a clear transition between two regimes, which is modeled using a segmented power-law fit. At low $\tilde{L}_a$ ($\tilde{L}_a \ll 1$), the data follow one scaling relation, while at high $\tilde{L}_a$ ($\tilde{L}_a \gg 1$), a distinct trend is observed.

Fitting the combined dataset for both progenitors (10 and 25~$M_\odot$) with a two-segment power-law model yields:
\begin{subequations}\label{eq:combined_fit}
\begin{align}
\textbf{Fit 1:} \quad 
N_s &= (0.957 \pm 0.145)\,\tilde{L}_a^{\,0.23 \pm 0.10}, 
\quad (\tilde{L}_a \gg 1), \label{eq:combined_fit_a} \\
\textbf{Fit 2:} \quad 
N_s &= (0.345 \pm 0.139)\,\tilde{L}_a^{\,0.25 \pm 0.13}, 
\quad (\tilde{L}_a \ll 1). \label{eq:combined_fit_b}
\end{align}
\end{subequations}

The fits yield $R^2 = 0.67$ (Fit~1) and $0.45$ (Fit~2).

% \begin{figure*}
% \centering

% \begin{subfigure}[t]{0.48\textwidth}
%     \includegraphics[width=\linewidth]{Fit_TwoSegment_Formatted_10Msun.pdf}
%     \caption{}
% \end{subfigure}
% \hfill
% \begin{subfigure}[t]{0.48\textwidth}
%     \includegraphics[width=\linewidth]{Fit_TwoSegment_Formatted_25M.pdf}
%     \caption{}
% \end{subfigure}

% \caption{Two-segment power-law fit of the calibration constant $N_s$ as a function of $\tilde{L}_a$. 
% \textbf{Panel (a):} Result for a $10\,M_{\odot}$ progenitor star. 
% \textbf{Panel (b):} Result for a $25\,M_{\odot}$ progenitor star. 
% Green markers represent the Newtonian regime, blue markers represent the relativistic regime, and the transition regime is indicated by the grey shaded region.}

% \label{fig:fit_two_segment}

% \end{figure*}

In the Newtonian regime, we obtain the empirical relation given by equation \ref{eq:combined_fit_b}. 
% Incorporating this calibration 
% into $\beta_h \approx \tilde{L}_s^{1/2}$ yields a scaling of $\beta_h \propto \tilde{L}_a^{0.75\pm 0.13}$. This, in turn, implies a breakout time dependence $t_B \propto L_j^{-0.75^{+0.44}_{-0.26}}$. Although the lower 
% bound of the predicted range marginally overlaps with the simulation results given by equation \ref{eq:10Msun_low} and equation \ref{eq:25Msun_low}, the central value and most of the allowed range remain 
% significantly steeper. {\bf Thus, even after incorporating the calibration factor, the analytical scaling systematically deviates from the simulation-derived relation and fails to robustly reproduce the observed trend.}
Incorporating the calibration, including both the normalization and the power-law index of the relation between $\tilde{L}_s$ and $\tilde{L}_a$, into $\beta_h \approx \tilde{L}_s^{1/2}$ yields a scaling of $\beta_h \propto \tilde{L}_a^{0.75\pm0.13}$. This, in turn, implies a breakout-time dependence of $t_B \propto L_{jet}^{-0.75^{+0.44}_{-0.26}}$. While the lower bound of the predicted power-law index marginally overlaps with the 
slopes obtained from the simulations, as given by equations \ref{eq:10Msun_low} and \ref{eq:25Msun_low}, the central value and the majority of the allowed range remain significantly steeper than the simulation-derived slopes. 
Moreover, the breakout times predicted by the calibrated analytical scaling do not reproduce the breakout times measured in the hydrodynamic simulations (Figure \ref{fig:breakout_calibrated}a). Thus, 
even after incorporating the calibration of both the normalization and the power-law index, the analytical model fails to reproduce both the $L_{jet}$-dependence and the absolute values of the breakout times obtained from the simulations.

\subsection{Analytical Relation: Calibration in the Relativistic Regime}
\label{calibration_relativistic}

\begin{figure*}[t]
\centering

\gridline{
\fig{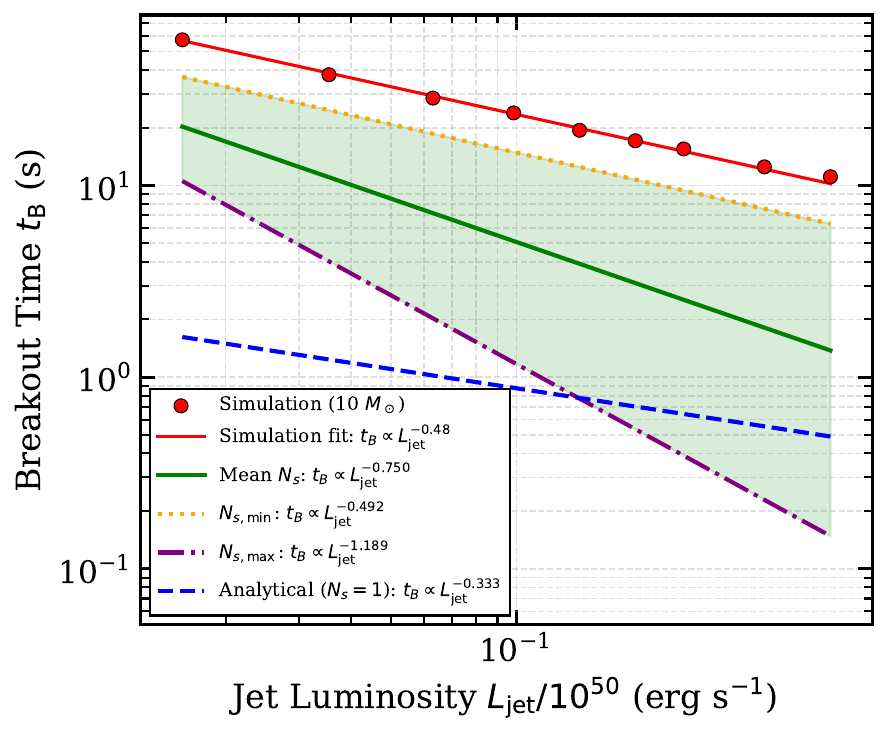}{0.45\textwidth}{(a)}
\fig{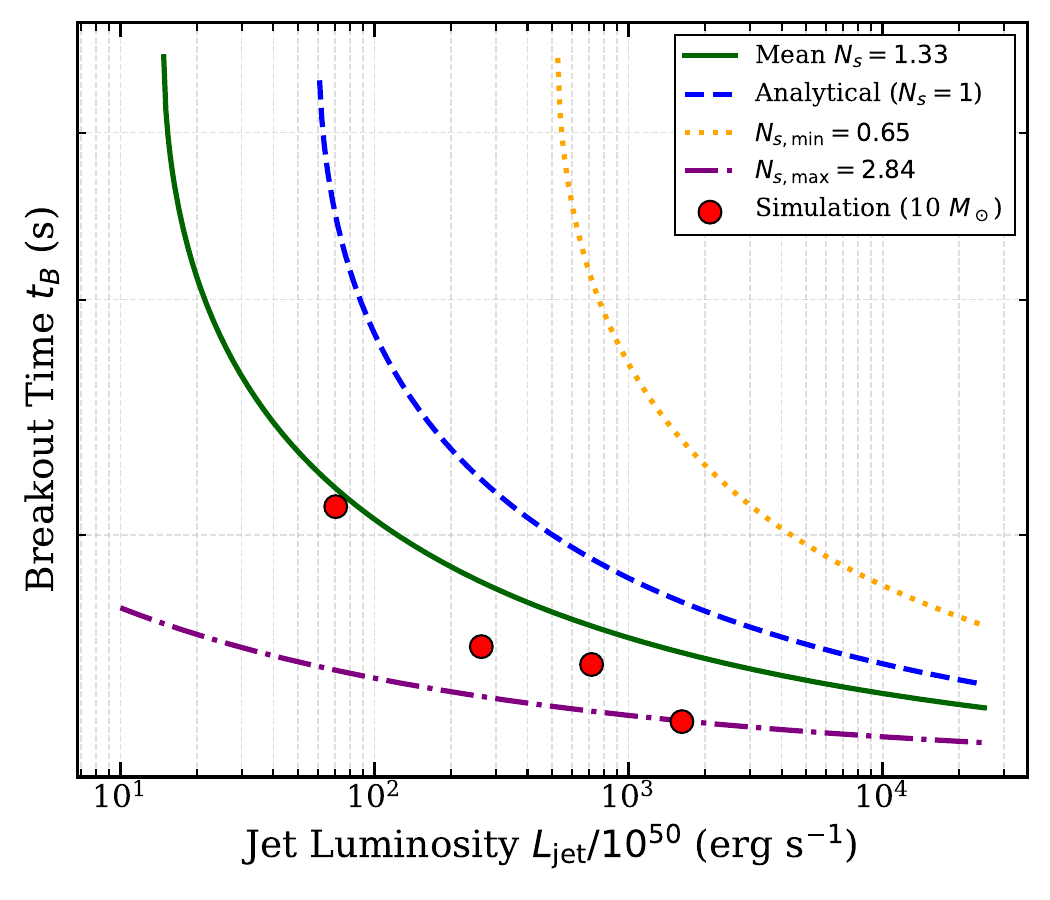}{0.45\textwidth}{(b)}
}
\caption{
% \textbf{(a)} Jet breakout time $t_B$ as a function of jet luminosity in the low-luminosity Newtonian regime. Blue squares denote the numerical simulation results, while the solid blue line represents the power-law fit to the simulations. The red dotted line shows the breakout-time relation obtained from the calibration analysis, while the shaded region represents the corresponding bounds arising from the uncertainties in the calibration.
% \textbf{(b)} Jet breakout time $t_B$ as a function of jet luminosity
% $L_j$ for the $10\,M_\odot$ progenitor. Red circles denote the numerical
% simulation results. The solid dark green line indicates the calibrated
% model using the mean calculated calibration constant, while the blue
% dashed line represents the standard analytical case ($N_s=1$). The
% orange dotted and purple dash-dotted curves correspond to the minimum
% and maximum $N_s$ values, respectively. The horizontal dotted line
% represents the light-travel limit $R/c$.
\textbf{(a)} Jet breakout time $t_B$ versus jet luminosity $L_{jet}$ for the $10\,M_\odot$ progenitor in the low-luminosity Newtonian regime. Red circles show the simulations and the solid red line shows their power-law fit. The solid dark green, blue dashed, orange dotted, and purple dash-dotted lines show the power-law relations from the calibration analysis for the mean, $N_s=1$, minimum, and maximum $N_s$, respectively, with the shaded region indicating the calibration uncertainty. \textbf{(b)} Same as (a), but for the relativistic regime, where the corresponding lines show the model curves.
}
\label{fig:breakout_calibrated}
\end{figure*}

% \begin{figure}
% \centering
% \fig{breakout_calibrated_10.pdf}{0.95\columnwidth}{}
% \caption{
% Jet breakout time $t_B$ as a function of luminosity $L_j$ for a
% $10\,M_\odot$ progenitor. Red circles denote results from numerical
% simulations. The solid dark green line indicates the calibrated model
% using the mean calculated calibration constant, while the blue dashed
% line represents the standard analytical case ($N_s = 1$). The orange
% dotted and purple dash-dotted curves correspond to the minimum and
% maximum $N_s$ values, respectively. The horizontal dotted line
% represents the light-travel limit $R/c$.
% }
% \label{fig:breakout_calibrated}
% \end{figure}

In the relativistic regime ($\tilde{L} \gg 1$), the jet-head velocity approaches the speed of light and can be approximated as \citep{Bromberg_11}

\begin{equation}
\beta_h = \frac{1}{1 + \tilde{L}^{-1/2}} \approx 1 - \frac{\tilde{L}^{-1/2}}{2}.
\label{eq:rel_beta}
\end{equation}

Substituting into the jet propagation condition,

\begin{equation}
R = c \int_0^{t_B} \beta_h(t)\, dt,
\end{equation}

% we obtain

% \begin{equation}
% R = c \int_0^{t_B} \left(1 - \frac{\tilde{L}^{-1/2}}{2} \right) dt,
% \end{equation}

% which gives

% \begin{equation}
% \frac{R}{c} = t_B - \int_0^{t_B} \frac{\tilde{L}^{-1/2}}{2} dt.
% \end{equation}

%Rearranging,
we obtain, 

\begin{equation}
t_B = \frac{R}{c} + \int_0^{t_B} \tilde{L}^{-1/2} dt.
\label{eq:tB_rel_int}
\end{equation}

Using the analytical scaling of the dimensionless luminosity in the collimated regime,

\begin{equation}
\tilde{L}^{-1/2} \propto L_{jet}^{-1/5} \, t^{2/5},
\end{equation}

we obtain

\begin{equation}
t_B = \frac{R}{c} + k \, L_{jet}^{-1/5} \int_0^{t_B} t^{2/5} dt,
\end{equation}

% Evaluating the integral,

% \begin{equation}
% \int_0^{t_B} t^{2/5} dt = \frac{5}{7} t_B^{7/5},
% \end{equation}

% we finally obtain
which, on further evaluation of the integral, yields,

\begin{equation}
t_B = \frac{R}{c} + k \, L_{jet}^{-1/5} \, t_B^{7/5},
\label{eq:tB_rel_final}
\end{equation}

where $k$ is a constant that depends on the ambient density profile and jet properties, defined as:
\[
k = \frac{5}{7} \left( \frac{3 \pi}{16 \Omega } \theta_0^4 \rho_a c^5 \right)^{1/5}.
\label{eq:k_definition}
\]
%Here, $\Omega$ is the dimensionless integration parameter, $\rho$ is the density of star, and $\theta_{0}$ is the opening angle.
% where $k$ is a constant that depends on the ambient density profile and jet properties.

%\bigskip

Equation~\ref{eq:tB_rel_final} shows that, in the relativistic regime, the breakout time is primarily set by the light-crossing time $R/c$, with only a weak luminosity-dependent correction. In contrast to the Newtonian 
regime, which exhibits a relatively stronger power-law dependence, the relativistic case shows a much weaker sensitivity to jet luminosity. A comparison between the numerical simulations and the analytical 
predictions is presented in Figure~\ref{fig:breakout_calibrated}b. We find that the standard analytical choice of the integration parameter $\Omega = 1.47$ (dashed blue line; see Appendix A1 of \harrisonEighteen) systematically overestimates the breakout time across all luminosities for both progenitors.

To address this discrepancy, we include the calibration factor $N_s$, defined through the relation between the simulated and analytical values of $\tilde{L}$ (Equation~\ref{eq:Ns}) in the relativistic regime. Incorporating this calibration yields
\begin{equation}
t_B = \frac{R}{c} + \frac{k}{N_s} \, L_{jet}^{-1/5} \, t_B^{7/5},
\label{eq:tB_calibrated_final}
\end{equation}

Using the average calibration factor obtained in the relativistic regime ($a > 0.2$), $N_s \approx 1.32$ for the $10\,M_{\odot}$ progenitor, the calibrated model (green solid line) provides a significantly improved match to the simulation data (Figure~\ref{fig:breakout_calibrated}b). To account for the variation of $N_s$ with $L_{jet}$, we also show curves corresponding to the minimum (dash-dotted purple) and maximum (dotted yellow) $N_s$ values. The corresponding calibration and breakout-time comparison for the $25\,M_{\odot}$ progenitor are presented in Appendix~\ref{sec:appendix_25Msun_calibration}.

% The slightly higher $N_s$ required for the $25\,M_\odot$ star is consistent with the stronger jet--cocoon coupling expected in denser, higher-mass environments. Consequently, $N_s$ serves as a robust physical bridge between idealized steady-state theory and the non-linear dynamics of numerical simulations.
%%%%%%

% \begin{figure}[H]
% \centering
% \includegraphics[width=0.48\textwidth]{Relativistic regime- Analytical solution/breakout_calibrated_envelope_10.pdf}
% \includegraphics[width=0.48\textwidth]{Relativistic regime- Analytical solution/breakout_calibrated_envelope_25.pdf}
% \caption{Jet breakout time $t_B$ as a function of luminosity $L_j$ for $10\,M_\odot$ (left) and $25\,M_\odot$ (right) stellar progenitors. Red circles denote simulation data points, while the solid navy lines represent the best-fit model with optimized $\Omega$ parameters. The shaded blue regions indicate the 1-$\sigma$ uncertainty of the fits. For comparison, the dashed orange lines show the analytical solution ($\Omega = 1.47$). The horizontal dotted lines represent the light-travel limit $R/c$. }
% \label{fig:relativistic_regime_solution}
% \end{figure}

\section{Discussion}
\label{Discussion}
\subsection{Jet Breakout and Choking}
\label{Dis_breakout}
\begin{figure*}
\centering

\gridline{
\fig{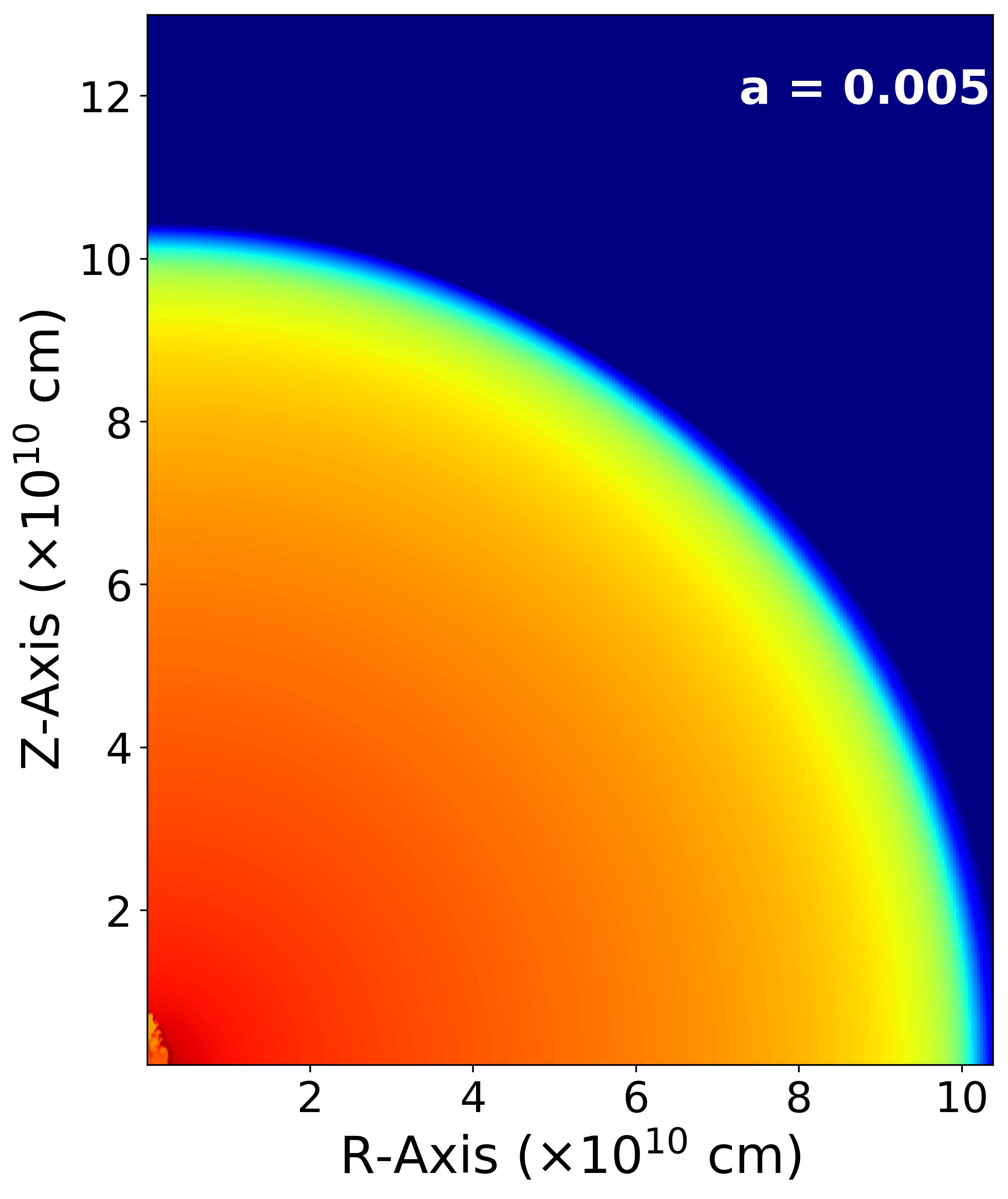}{0.308\textwidth}{(a)}
\fig{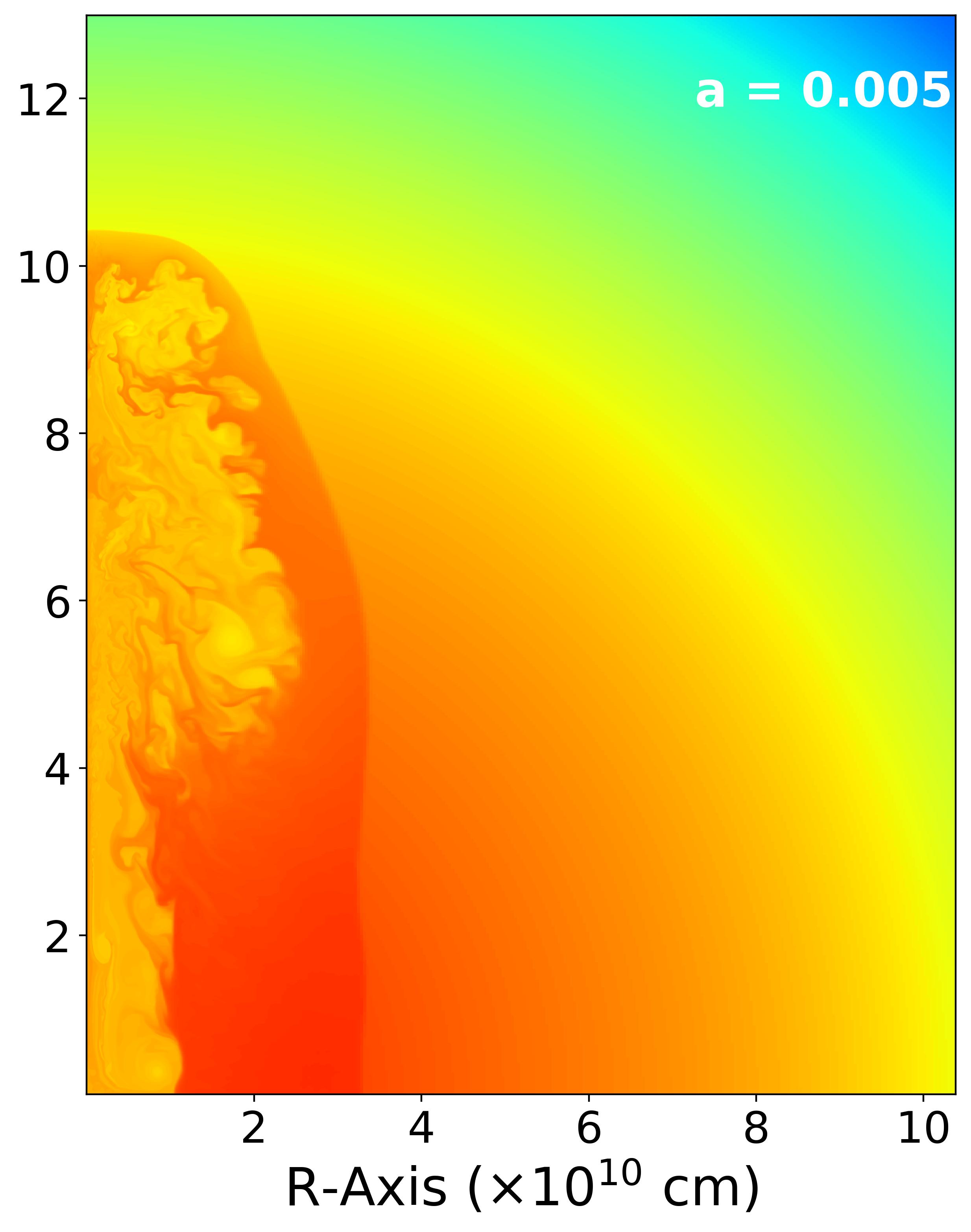}{0.29\textwidth}{(b)}
\fig{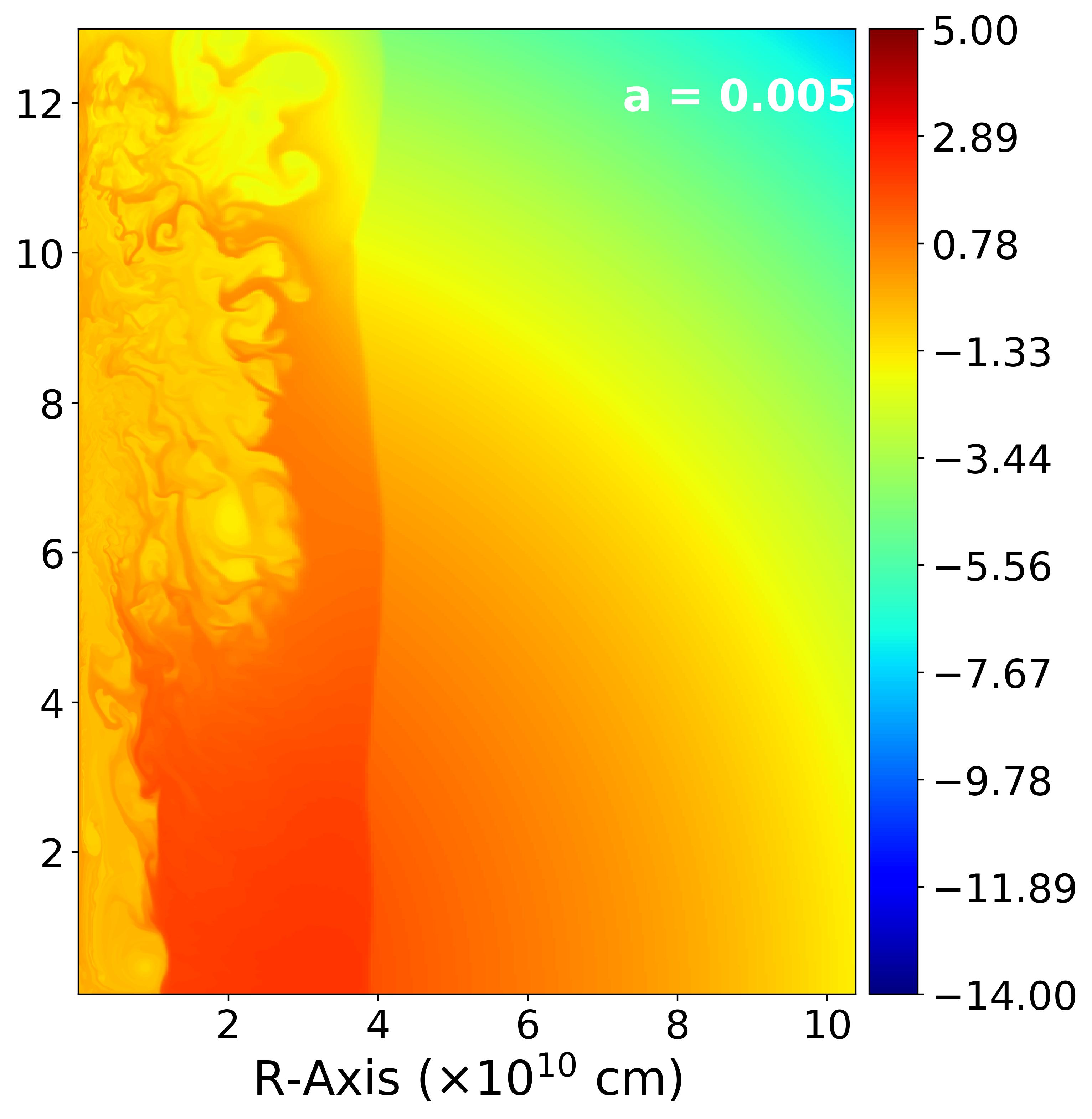}{0.36\textwidth}{(c)}
}

\caption{
Logarithmic density maps of the stellar envelope for a $25\,M_{\odot}$ progenitor with a black hole spin of $a = 0.005$ at three different times. 
From left to right, panels (a), (b), and (c) correspond to $t = 20$ s, $t = 320$ s, and $t = 380$ s. 
The sequence illustrates the temporal evolution of the stellar envelope as it expands and becomes progressively restructured with time.
}  

\label{fig:Envelope_25}

\end{figure*}

We study the dependence of jet breakout on black hole spin and determine the threshold below which the jet becomes choked. Our simulations show that for very low spins ($a\leq 0.001$) the jet fails to break out of the star within 
the fallback timescale. This is consistent with earlier numerical studies such as those by \cite{Bromberg_11b, Aloy_2018, Mizuta_and_Ioka_2013}, who demonstrated that jets with insufficient luminosity are unable to overcome the ram pressure of the stellar envelope and become choked before breakout.
In the $10\,M_{\sun}$ progenitor, the jet stalls well inside the star and remains confined there until the free-fall time. However, in the 
$25\,M_{\sun}$ progenitor, we observe that, as the simulation approaches the free-fall time, the outer stellar envelope is gradually displaced outward from its initial position (Figure \ref{fig:Envelope_25}). We also find that in both the $10 M_{\sun}$ and $25M_{\sun}$ progenitors, when the jet injection luminosity is significantly lower (corresponding to $a < 0.01$), the outer stellar envelope tends to be uplifted and displaced outward (Figure \ref{fig:ChokedJets} and Figure \ref{fig:Envelope_25}).

We also examine the jet-head velocities during these phases. In the $10\,M_{\sun}$ ($25\,M_{\sun}$) progenitor, the jet head reaches velocities of $\beta_h \approx 0.4-0.9$ ($0.3-0.9$) at the time of breakout, whereas in the stalled cases, the jet-head velocity remains around $\beta_h\approx0.007$, well below the $\beta_h\sim0.1$ choking threshold reported by \cite{Hamidani_2025}, consistent with their finding that jets with $\beta_h\lesssim0.1$ remain choked. These jet-head velocities at successful breakout are consistent with previous numerical simulation studies (e.g., \citealt{Hamidani_2025, Gottlieb_2022}), which report that successful jets attain mildly relativistic to relativistic speeds near the stellar surface. This indicates that once the jet head attains relativistic velocities, it is 
able to successfully penetrate the stellar envelope, while lower velocities correspond to stalled or choked jets. In the $25\,M_{\sun}$ progenitor, we additionally observe that the jet begins to expand more laterally as it approaches threshold spin values and eventually emerges in a more diffuse manner after the outer stellar envelope is pushed outward (Figure \ref{fig:Envelope_25}).

% \label{Dis_breakout}
% \begin{figure*}
%     \centering
%     % First image
%     \begin{subfigure}[H]{0.308\textwidth}
%         \includegraphics[width=\linewidth]{rho_spin_0.005_20_25.jpg}
%         %\caption{}  % optional caption
%     \end{subfigure}
%     \hfill
%     % Second image
%     \begin{subfigure}[H]{0.29\textwidth}
%         \includegraphics[width=\linewidth]{rho_spin_0.005_320_25.jpg}
%         %\caption{}  % optional caption
%     \end{subfigure}
%     \hfill
%     % Third image
%     \begin{subfigure}[H]{0.36\textwidth}
%         \includegraphics[width=\linewidth]{rho_spin_0.005_380_25.jpg}
%         %\caption{}  % optional caption
%     \end{subfigure}

%     % Overall figure caption
%    \caption{Logarithmic density maps of the stellar envelope for a $25\,M_{\odot}$ progenitor with a black hole spin of $a = 0.005$ at three different times. From left to right, the snapshots correspond to $t = 20$ s, $t = 320$ s, and $t = 380$ s. The sequence illustrates the temporal evolution of the stellar envelope as it expands and becomes progressively restructured with time.}

%     \label{fig:Envelope_25}
% \end{figure*}

The outward displacement of the envelope likely arises from the inflation of a hot cocoon composed of shocked jet and stellar material surrounding the jet. As the jet 
propagates, energy deposited into the cocoon can drive lateral expansion and exert pressure on the surrounding stellar layers, thereby modifying the density structure of the stellar environment. This effect is particularly 
pronounced in the $25\,M_{\sun}$ progenitor, where the longer free-fall timescale allows the simulation to run for a longer duration. In the low-spin case ($a=0.005$), the 
extended runtime likely results in additional cocoon heating, causing it to expand further and push the outer envelope outward.

Although the black hole spin is kept constant in our simulations, the behaviour of the jet for different BH spins obtained in this study may provide insight into a more realistic scenario in which the BH spin 
evolves with time in a GRB event. In particular, a gradual spin-down of the black hole could reduce the jet power, eventually leading to jet choking and the termination of the prompt GRB emission. If the displaced envelope material subsequently 
falls back toward the central compact object, the resulting fallback accretion may temporarily revive the central engine and produce renewed jet activity after a quiescent phase. Such a mechanism could potentially 
manifest as late-time flaring or episodic emission. A detailed investigation of this scenario requires modelling the time-dependent evolution of the central engine and fallback accretion, which we plan to explore in future work.

\subsection{Comparison with Previous Calibrations}
\label{comparison_calibrations}

The calibration obtained in this work is compared with previous studies in Table~\ref{tab:calibration_comparison}. Unlike previous studies that primarily varied the ambient density structure or jet opening angle, we vary the jet luminosity while keeping the stellar density profile and jet injection geometry fixed. This allows us to directly examine the dependence of the correction factor $N_s$ on jet luminosity.

In the Newtonian regime, our calibration gives lower $N_s$ values ($\simeq 0.11 - 0.27$) than the $N_s\simeq0.3-0.4$ found by \harrisonEighteen . This difference is expected given the different parameter spaces: \harrisonEighteen{} varied the stellar density structure at nearly fixed jet luminosity, whereas we vary the luminosity for a fixed stellar structure. A similar dependence on the physical setup was noted by \citet{Hamidani_2021}, who found $N_s\simeq0.3-0.4$ for collapsar models but substantially different values for BNS-merger models. Their Appendix~C explicitly shows that $N_s$ depends on the physical scenario
and the parameter space considered. In particular, their collapsar
models give $N_s\simeq0.3-0.4$, whereas substantially different values
are obtained for their BNS-merger models. They further caution that
the calibrated values of $N_s$ should be associated with the parameter
space in which they are derived. This is particularly relevant here,
as our Newtonian simulations probe $\tilde{L}_a$ values below the
$\tilde{L}\sim0.1-1$ range emphasized in their calibration. The lower
$N_s$ values found in our simulations can therefore be understood as a
luminosity-dependent calibration in a different region of parameter
space, rather than as a direct contradiction of their result.

In the transition regime, our $N_s$ values overlap with those of \harrisonEighteen, although the physical setups and $\tilde{L}_a$ ranges differ. In the relativistic regime, our results show a systematic dependence of $N_s$ on $\tilde{L}_a$, in contrast to the approximation $N_s\simeq1$ adopted by \harrisonEighteen. Overall, these comparisons indicate that $N_s$ is not necessarily a universal constant, but depends on the jet and ambient-medium properties. Our luminosity-dependent calibration is therefore most appropriate for the parameter space explored here and can be used to derive the corresponding luminosity dependence of the jet-head velocity and breakout time.

\begin{table*}[!t]
\centering
\caption{Comparison of the calibration of the jet-head velocity with previous works.
The different physical scenarios considered in each study are listed along with the
corresponding values of $N_s$ and ranges of $\tilde{L}_a$ in the Newtonian, transition,
and relativistic regimes.}
\label{tab:calibration_comparison}
\begin{tabular}{l l cc cc cc}
\hline\hline
\multirow{2}{*}{Physical scenarios} &
\multirow{2}{*}{Reference} &
\multicolumn{2}{c}{Newtonian} &
\multicolumn{2}{c}{Transition} &
\multicolumn{2}{c}{Relativistic} \\
\cline{3-8}
& &
$N_s$ & $\tilde{L}_a$ range &
$N_s$ & $\tilde{L}_a$ range &
$N_s$ & $\tilde{L}_a$ range \\
\hline

 Jet power
& This work
& $\propto \tilde{L}_a^{{\,0.25 \pm 0.13}}$ & $0.008-0.05$
& $0.20-0.50$ & $0.06-0.95$
& $\propto \tilde{L}_a^{\,0.23 \pm 0.10}$ & $2.7-12.5$ \\

Density medium
& \citealp{Harrison_2018}
& $0.3-0.4$ & $0.01 - 1$
& $0.35-1$ & $1-100$
& $1$ & $>100$ \\

% Changing circumburst medium
% & Hamidani et al. (2025)
% & -- & --
% & -- & --
% & -- & -- \\

Jet opening angle
& \citealp{Hamidani_2021}
& 0.38 & $0.1-1$
& -- & --
& -- & -- \\
\hline\hline
\end{tabular}
\end{table*}

\subsection{Jet Breakout Dynamics: Analytical Expectations and Simulation Results - Newtonian Regime}
\label{Dis_Newtonian}

The propagation speed of the jet head is determined by the balance between the ram pressure of the jet and that of the ambient stellar medium in the jet-head frame. In the 
analytical treatment of jet propagation by \citealt{Matzner_2003}, this balance is considered under the simplifying assumption that the jet opening angle and cross-sectional area remain constant. Under these conditions, the jet-head velocity 
scales approximately as $\beta_h \propto \tilde{L}^{1/2}$.
%where $\tilde{L}$ represents the ratio of the jet energy density to that of the surrounding medium. 
For collimated 
jets propagating through a stellar envelope, subsequent analytical work has shown that this relation leads to the breakout-time scaling $t_B \propto L_{\rm jet}^{-1/3}$ (e.g., \citealt{Bromberg_11}). Breakout time decreases with increasing jet luminosity, implying that the jet-head velocity must increase as $L_{\rm jet}$ increases, since 
$t_B \sim \frac{R_*}{\beta_h c}$.  

However, our simulations show that in the Newtonian regime the breakout time scales approximately as $t_B \propto L_{\rm jet}^{-0.48}$, which is steeper than the analytical expectation. This implies that higher-luminosity jets 
escape the stellar envelope more rapidly than predicted by the simple analytic model. 
%%%To understand this discrepancy, we examined how the jet-head velocity scales with $\tilde{L}$ in our simulations. 
 Furthermore, in section \ref{Newtonian_calibration}, we show that despite including the calibration factor $N_s$ into the analytical estimate does not resolve this discrepancy. 

This suggests that the discrepancy does not arise simply from the geometric effects associated with the evolution of the jet cross-section, as these are already accounted for through the calibration factor $N_s$.
Instead, the remaining difference may indicate that the underlying assumption $\beta_h \propto \tilde{L}_s^{1/2}$ does not adequately describe the jet-head dynamics in the regime considered here. This scaling follows from the simple ram-pressure balance between the jet and the surrounding 
stellar medium, as discussed by \citealt{Matzner_2003}. In the Newtonian regime, however, the jet–ambient interaction can be considerably more complex, involving enhanced 
turbulence and shock formation, as also evident in our numerical simulations (Figure \ref{fig:rho_spins_combined_10}). Such effects can modify the jet-head dynamics and lead to deviations from the standard $\beta_h$–$\tilde{L}_s$ scaling, potentially explaining the discrepancy between the analytical prediction and the simulation-derived relation.

We, thereby, investigate this by  
%To investigate this discrepancy, we 
generalizing the jet-head velocity relation as

\begin{equation}
\beta_h \approx \tilde{L}_s^{\delta},
\label{eq:21}
\end{equation}

where $\delta$ is treated as an unknown parameter. Using the relation $\tilde{L}_s = \tilde{L}_a \, N_s^2$, this can be written as

\begin{equation}
\beta_h \approx \tilde{L}_a^{\delta} \, N_s^{2\delta}.
\label{eq:25}
\end{equation}

Let us consider a scaling between  $N_s$ and $ \tilde{L}_a$ as 

\begin{equation}
N_s \sim A \tilde{L}_a^{p},
\end{equation}

where, $A$ is normalisation and $p$ is the power law index, which in turn gives 
\begin{equation}
\beta_h \sim A^{2\delta} \tilde{L}_a^{(2p+1)\delta}.
\end{equation}

Combining this with the analytical expression for $\tilde{L}_a$ (Equation~\ref{eq:collimated}) and following the standard breakout time derivation, we obtain a generic relation between $t_B$ and $L_j$. The detailed generic numerical expression is given in Appendix \ref{generic_relation}. 

% \begin{equation}
% t_B \propto L_j^{\frac{2 (2p+1)\delta}{4 (2p+1)\delta - 5}}.
% \label{eq:22}
% \end{equation}

Matching this relation with the simulation-derived scaling 
$t_B \propto L_j^{-0.48}$ (Equation~\ref{eq:10Msun_low}) yields 
$(2p+1)\delta = 0.61$. For the $25\,M_{\odot}$ progenitor, using the corresponding scaling $t_B \propto L_j^{-0.53}$, we obtain $(2p+1)\delta \approx 0.64$. 

% The simulation results indicate that the jet breakout occurs earlier than predicted by the standard analytical model, implying that the jet head propagates faster than expected.  
% Since $\tilde{L}_s<1$ in the Newtonian regime, this suggests that the jet-head velocity may have a weaker dependence on $\tilde{L}_s$ than the standard 
% $\beta_h\propto\tilde{L}_s^{1/2}$ scaling. Specifically, an effective scaling of $\beta_h\propto\tilde{L}_s^\delta$ with 
% $\delta<1/2$ would yield higher $\beta_h$ values for a given $\tilde{L}_s$. Thus, in the Newtonian regime, using the relation between $\delta$ and $p$,

% \begin{equation}
% \frac{0.61}{2p+1} < \frac{1}{2},
% \label{detla_p_relation}
% \end{equation}

% gives

% \begin{equation}
% p > 0.11, 
% \end{equation}
% and comparing the normalisation of the simulation driven scaling between $t_B$ and $L_j$, we get the constrain that 
% \begin{equation}
% A < 0.158, 
% \end{equation}

The simulation results (Figure \ref{fig:breakout_calibrated}a) indicate that the jet breaks out later than predicted by the standard analytical model, implying a slower propagation of the jet head than expected. Since $\tilde{L}_s<1$ in 
the Newtonian regime, this discrepancy suggests that the jet-head velocity has a stronger dependence on $\tilde{L}_s$ than the standard scaling $\beta_h\propto\tilde{L}_s^{1/2}$. In particular, adopting 
an effective scaling of the form $\beta_h\propto\tilde{L}_s^\delta$, a value $\delta>1/2$ would result in a higher jet-head velocity for a given $\tilde{L}_s$. Using the relation between $\delta$ and $p$ in the Newtonian regime,
\begin{equation}
\delta = \frac{0.61}{2p+1},
\end{equation}
the requirement $\delta>1/2$ gives
\begin{equation}
\frac{0.61}{2p+1}>\frac{1}{2},
\label{detla_p_relation}
\end{equation}
which implies
\begin{equation}
p<0.11.
\end{equation}
Furthermore, comparison of the normalization of the simulation-derived $t_B$--$L_j$ relation with the corresponding analytical scaling places an upper constraint on the normalization parameter,
\begin{equation}
A>0.158.
\end{equation}

% It is interesting to note that, assuming $\delta=1/2$, the resulting scaling between $N_s$ and $\tilde{L}_a$ gives $0.12 < p < 0.38$ (i.e. equation \ref{eq:combined_fit_b}). This is inconsistent with the constraint $p<0.11$ inferred 
% from the shorter breakout times, providing further evidence that the standard $\delta=1/2$ scaling may not adequately describe the simulations.

% Furthermore, if $N_s$ is approximately independent of $\tilde{L}_a$, as reported by \citealt{Harrison_2018}, then 
% the corresponding relation yields $\delta=0.61$. This would further support a value of $\delta$ significantly larger than $1/2$. 

Overall, the deviation from the standard $\beta_h$–$\tilde{L}_s$ scaling may arise from the strong interaction between the jet and the surrounding stellar material in the Newtonian regime. As evident from Figure \ref{fig:rho_spins_combined_10}a,b,c, this interaction 
generates shocks and turbulent structures that promote substantial mixing between the jet and stellar material. Meanwhile, the lateral expansion of the cocoon exerts pressure on the jet, enhancing the entrainment of stellar 
material into the jet head and increasing its effective inertia. The resulting redistribution and dissipation of momentum reduce the net jet-head propagation velocity relative to that predicted by the standard analytical scaling, leading to a systematically longer breakout time. 

\subsection{Jet Breakout Dynamics: Analytical Expectations and Simulation Results - Relativistic Regime}
\label{Dis_Relativistic}
GRBs are powered by relativistic jets launched from accreting black holes, with the large observed energetics commonly attributed to the extraction of rotational energy of BH via magnetic fields threading the black hole through the Blandford–Znajek mechanism. In 
this work, we consider a scenario in which the black hole spin remains approximately constant during both the jet breakout phase and the burst duration.

Previous studies \citep{Gottlieb_2023,Jacquemin-Ide_2024} have shown that, for typical GRB energetics and reasonable 
radiation efficiencies, the required jet luminosities are $\lesssim 10^{51}\,{\rm erg\,s^{-1}}$, corresponding to relatively low black hole spins ($a \lesssim 0.1$). 
This suggests that either GRBs are powered by relatively low spinning black holes or that initially high spins rapidly decrease due to energy extraction through the jet. Consequently, sustained high-spin configurations are not generally expected for the majority of GRBs.

However, a subset of particularly energetic events presents a different picture. For example, \citet{Sharma_etal_2021} identify several GRBs with 
total jet energies, $E_{\rm jet}$ exceeding $10^{52}\,{\rm erg}$. For these bursts, the inferred source-frame jet luminosities, estimated as $L_{\rm jet} = E_{\rm jet}
(1+z)/T_{90}$, where $z$ is the redshift and $T_{90}$ corresponds to the duration of the burst, are typically below $10^{51}\,{\rm erg\,s^{-1}}$, but a few notable cases such as GRB~110731A and GRB~090926A reach values of $\sim 4.5\times10^{52}\,{\rm erg\,s^{-1}}$ and $\sim 
7.5\times10^{51}\,{\rm erg\,s^{-1}}$, respectively. Such luminosities correspond to moderately high black hole spins in the range $a \sim 0.2$--$0.3$, assuming 
approximately steady energy injection. Integrating these luminosities over the burst duration naturally explains the large total jet energies inferred for these events.

These rare, high-luminosity GRBs may, therefore, be powered by moderately high-spin black holes operating in the relativistic regime with sustained jet activity. In contrast, the bulk of the GRB population is likely associated with lower spins or with systems in which the black hole undergoes significant spin-down during the burst, as expected in magnetically arrested accretion 
scenarios. Sustaining the high luminosities observed in the brightest events requires that the jet power remain nearly constant over a substantial fraction of the burst duration. This, in turn, implies that the black hole spin must 
remain approximately constant or evolve slowly, which could occur if the angular momentum supplied by accretion balances, or slightly exceeds, that extracted through the Blandford–Znajek mechanism.

Motivated by this, we investigate the dependence of the jet breakout time on $L_{\rm jet}$ in the relativistic regime under the assumption of constant spin, and examine its implications for jet propagation (Section~\ref{calibration_relativistic}). From Equation~\ref{eq:tB_rel_final}, the breakout time is expected to exceed $R_s/c$ due to an additional correction term that scales as $L_{jet}^{-1/5}$. For the $10\,M_{\odot}$ and $25\,M_{\odot}$ progenitors, $R_s/c$ is 1.33\,s and 3.33\,s, respectively. As the spin $a$ (and hence $L_{jet}$) increases, the contribution of this correction term diminishes, causing $t_B$ to asymptotically approach the minimum timescale $R_s/c$. 

We find that the analytical model systematically overestimates the breakout time for a given jet luminosity in the relativistic regime. Incorporating the calibration factor $N_s$ significantly improves the agreement with simulations; however, even after calibration (using the exact values obtained in Figure \ref{fig:calibration}), the simulation data points tend to lie toward the lower bound of the predicted range. The systematic tendency of the simulation data to lie toward the lower bound of the predicted range suggests that the 
analytical model slightly overestimates the resistance to jet propagation. This likely arises from simplifying assumptions such as steady-state propagation, idealized density profiles, and efficient 
cocoon confinement. In contrast, the simulations capture time-dependent acceleration of the jet head, multidimensional effects, and variations in the ambient medium, all of which can facilitate faster jet propagation and hence shorter breakout times. 

\section{Summary}
\label{Summary}

% In this work, we performed relativistic hydrodynamic simulations to investigate the conditions for successful and choked jets in two different progenitor stars. Our results indicate that a minimum black hole spin of $a \gtrsim 0.001$ is required for the jet to successfully break out of the stellar envelope. We explored a wide range of spin values ($0.001 \leq a \leq 0.9$) and identified two distinct regimes: a Newtonian regime at low spin ($a \lesssim 0.03$) and a relativistic regime at high spin ($a \gtrsim 0.2$).  We further examined the behavior in these regimes and found that, in both cases, the standard analytical model deviates from our simulation results.  In the Newtonian regime, we first attempted to calibrate the standard analytical model; however, the results were not consistent with our simulations. This discrepancy led us to examine the underlying scaling, where we found that the jet head velocity exhibits a weaker dependence on $\tilde{L}$ compared to analytical predictions. Incorporating this modified dependence into the analytical framework yields good agreement with the simulations. In contrast, in the relativistic regime, a simple calibration of the analytical model is sufficient, and the resulting predictions are consistent with our simulation results.

In this work, we perform a suite of two-dimensional relativistic hydrodynamic simulations to investigate jet 
propagation, breakout conditions, and choking in Wolf–Rayet progenitors of $10\,M_{\odot}$ and $25\,M_{\odot}$. Unlike previous studies that assume 
arbitrary jet injection, our approach is semi-analytical: the jet power is self-consistently linked to the black hole spin through an empirical relation 
motivated by GRMHD simulations. This enables controlled exploration of jet energetics across a wide luminosity range ($10^{48}$–$10^{53}\,\mathrm{erg\,s^{-1}}$) corresponding to spins $0.001 \leq a \leq 0.9$.

We determine the spin threshold for successful jet breakout, finding that jets are choked for $a \leq 0.001$, while $a > 0.001$ leads to successful breakout. This simulation suite allows us to establish a clear correlation between jet breakout time ($t_B$), 
jet luminosity ($L_{\rm jet}$), and black hole spin ($a$), revealing three distinct regimes: Newtonian ($a \lesssim 0.03$), relativistic ($a \gtrsim 0.3$), and an intermediate transition regime. Consistently, the jet head velocity at breakout exhibits a clear 
dichotomy: for $a \gtrsim 0.3$ ($\tilde{L} > 1$), the jet head reaches highly relativistic speeds with $\beta_h \gtrsim 0.8$, whereas for lower spins ($a < 0.2$), $\beta_h$ remains in the range $0.35$ - $0.8$.

% In the Newtonian regime, standard analytical models fail to reproduce the simulation trends. While analytical expectations predict $\beta_h \propto \tilde{L}^{1/2}$ and $t_B \propto L_{\rm jet}^{-1/3}$, our simulations yield a steeper dependence, $t_B \propto L_{\rm jet}^{-0.48}$. This discrepancy 
% reflects a weaker dependence of jet head velocity on $\tilde{L}$ and enhanced cocoon-driven collimation that modifies the effective jet cross-section. Incorporating this revised scaling into the analytical framework leads to good agreement with the simulations.

In the Newtonian regime, standard analytical models do not reproduce the trends found in our simulations. While the analytical scaling predicts $\beta_h \propto \tilde{L}^{1/2}$ and $t_B \propto L_{\rm jet}^{-1/3}$, we find a steeper luminosity dependence, $t_B \propto L_{\rm jet}^{-0.48}$, with systematically longer breakout times. This deviation likely results from the complex jet–cocoon interaction in our simulations, where dynamical collimation and mixing at the jet–cocoon interface modify the effective jet cross-section and momentum transport. The resulting time-dependent jet-head dynamics differ from the simplified assumptions adopted in the standard analytical model. Our simulations constrain the corresponding scaling parameters to $\delta > 1/2$, $p < 0.11$, and $A > 0.158$, providing quantitative bounds on the deviation from the standard analytical scaling.

% This deviation reflects modified jet-head dynamics arising from strong turbulence, mixing, and cocoon-driven collimation, which alter the effective jet cross-section 
% and momentum transport \textcolor{red}{we don't have viscous terms in the momentum equation? Numerical viscosity is there, but is it strong enough? is our flow turbulent? Obviously, in reality, a GRB jet will be turbulent!}. 

In the relativistic regime, the analytical model systematically overestimates the breakout time. Introducing a calibration factor $N_s$, derived from 
the relation between simulated and analytical $\tilde{L}$ considering the standard jet head scaling, significantly improves the agreement. However, even after calibration, the simulation 
results tend to lie toward the lower bound of the predicted range, indicating that the analytical model slightly overestimates the resistance to jet propagation. This likely arises from simplifying 
assumptions such as steady-state propagation and idealized density profiles, whereas the simulations capture time-dependent acceleration, multidimensional effects, and ambient medium variations that enable efficient jet propagation in terms of breakout time.

Overall, this work establishes a physically motivated, spin-dependent framework for jet energetics and breakout, and highlights the limitations of standard 
analytical models in capturing jet dynamics across different propagation regimes.

\section{Acknowledgment}
We thank Dr. Ore Gottlieb for his valuable contributions regarding the PLUTO code setup, helpful discussions on the jet propagation and the different highlighted regimes, and overall support throughout this work. We thank Prof. Dr. Christian Fendt for his valuable suggestions regarding the PLUTO code framework. S.I. is supported by DST INSPIRE Faculty Scheme (IFA19-PH245) and ANRF ARG Grant (ANRF/ARG/2025/000380/PS). We also acknowledge the High-Performance Computing facility, Padmanabha Cluster, of IISER TVM for providing the necessary computational resources and support for technical issues.

\section{DATA availability}
The simulation data underlying this article will be shared upon reasonable request with the corresponding authors.

\appendix
\section{Results for \texorpdfstring{$25\,M_{\odot}$}{25 Msun} Progenitor }
\subsection{Jet Breakout and Choking }
\label{sec:appendix_25Msun}

\begin{table*}
\centering
\renewcommand{\arraystretch}{1.2}
\setlength{\tabcolsep}{7pt}
\small
\begin{tabular}{ccccccc}
\hline
\hline
\textbf{Mass} & \textbf{Luminosity ($L_{jet}$)} & $\tilde{L}_a$ & $\tilde{L}_s$ & \textbf{Spin (a)} & \textbf{$t_B$} & $\boldsymbol{\beta_h}$ \\
($M_{\odot}$) & (erg/s) &  &  &  & (s) & (at breakout) \\
\hline

\multicolumn{7}{c}{\textbf{$10\,M_{\odot}$}} \\
\hline
10 & $1.6213\times10^{53}$ & 12.5813 & 49.3906 & 0.9000 & 1.45 & 0.845 \\
10 & $7.1516\times10^{52}$ & 7.2407  & 12.4076 & 0.7000 & 1.60 & 0.93 \\
10 & $2.6324\times10^{52}$ & 3.8001  & 6.2698  & 0.5000 & 1.65 & 0.831 \\
10 & $7.0373\times10^{51}$ & 2.7200  & 2.9276  & 0.3000 & 2.10 & 0.838 \\
10 & $2.7889\times10^{51}$ & 0.9535  & 0.2542  & 0.2000 & 5.00 & 0.890 \\
10 & $6.4640\times10^{50}$ & 0.3535  & 0.0396  & 0.1000 & 6.10 & 0.6 14 \\
10 & $2.2880\times10^{50}$ & 0.1492  & 0.0232  & 0.0600 & 11.10 & 0.621 \\
10 & $5.6789\times10^{49}$ & 0.0675  & 0.0013  & 0.0300 & 12.50 & 0.599 \\
10 & $3.9407\times10^{49}$ & 0.0516  & 0.0016  & 0.0250 & 15.50 & 0.549 \\
10 & $1.9293\times10^{49}$ & 0.0420  & 0.0007  & 0.0175 & 17.10 & 0.520 \\
10 & $1.4171\times10^{49}$ & 0.0293  & 0.0004  & 0.0150 & 19.40 & 0.769 \\
10 & $9.8395\times10^{48}$ & 0.0240  & 0.0003  & 0.0125 & 23.90 & 0.491 \\
10 & $3.5413\times10^{48}$ & 0.0190  & 0.0003  & 0.0075 & 37.80 & 0.419 \\
10 & $1.5738\times10^{48}$ & 0.0088  & 0.0001  & 0.0050 & 57.50 & 0.66 \\

\hline
\multicolumn{7}{c}{\textbf{$25\,M_{\odot}$}} \\
\hline
25 & $1.14\times10^{53}$ & 8.5692 & 28.9765 & 0.9000 & 3.80 & 0.950 \\
25 & $5.05\times10^{52}$ & 5.9558 & 14.0912 & 0.7000 & 4.10 & 0.864 \\
25 & $1.86\times10^{52}$ & 3.6040 & 7.6646  & 0.5000 & 4.20 & 0.950 \\
25 & $4.97\times10^{51}$ & 1.8020 & 2.1971  & 0.3000 & 4.40 & 0.950 \\
25 & $1.97\times10^{51}$ & 0.8756 & 0.1812  & 0.2000 & 6.40 & 0.456 \\
25 & $4.57\times10^{50}$ & 0.3742 & 0.0748  & 0.1000 & 11.60 & 0.377 \\
25 & $1.12\times10^{50}$ & 0.1901 & 0.0053  & 0.0500 & 16.50 & 0.581 \\
25 & $4.01\times10^{49}$ & 0.0935 & 0.0037  & 0.0300 & 24.30 & 0.440 \\
25 & $2.78\times10^{49}$ & 0.0483 & 0.0036  & 0.0250 & 28.90 & 0.613 \\
25 & $1.36\times10^{49}$ & 0.0436 & 0.0011  & 0.0175 & 35.80 & 0.492 \\
25 & $1.00\times10^{49}$ & 0.0271 & 0.0007  & 0.0150 & 41.80 & 0.534 \\
25 & $ 6.97\times10^{49}$ & 0.0209 & 0.0006  & 0.0125 & 47.80 & 0.591 \\
25 & $2.50\times10^{48}$ & 0.0084 & 0.0001  & 0.0075 & 91.30 & 0.478 \\

\hline
\end{tabular}
\caption{Comparison of analytical ($\tilde{L}_a$) and simulation-derived ($\tilde{L}_s$) dimensionless jet parameter $\tilde{L}$ for $10\,M_{\odot}$ and $25\,M_{\odot}$ progenitors. The table also lists the corresponding jet breakout time ($t_B$) and the jet head velocity at breakout ($\beta_h = v_h/c$) for different jet luminosities and black hole spin parameters.}
\label{tab:combined}
\end{table*}

\begin{figure*}
\centering

% -------- Top row --------
\gridline{
\fig{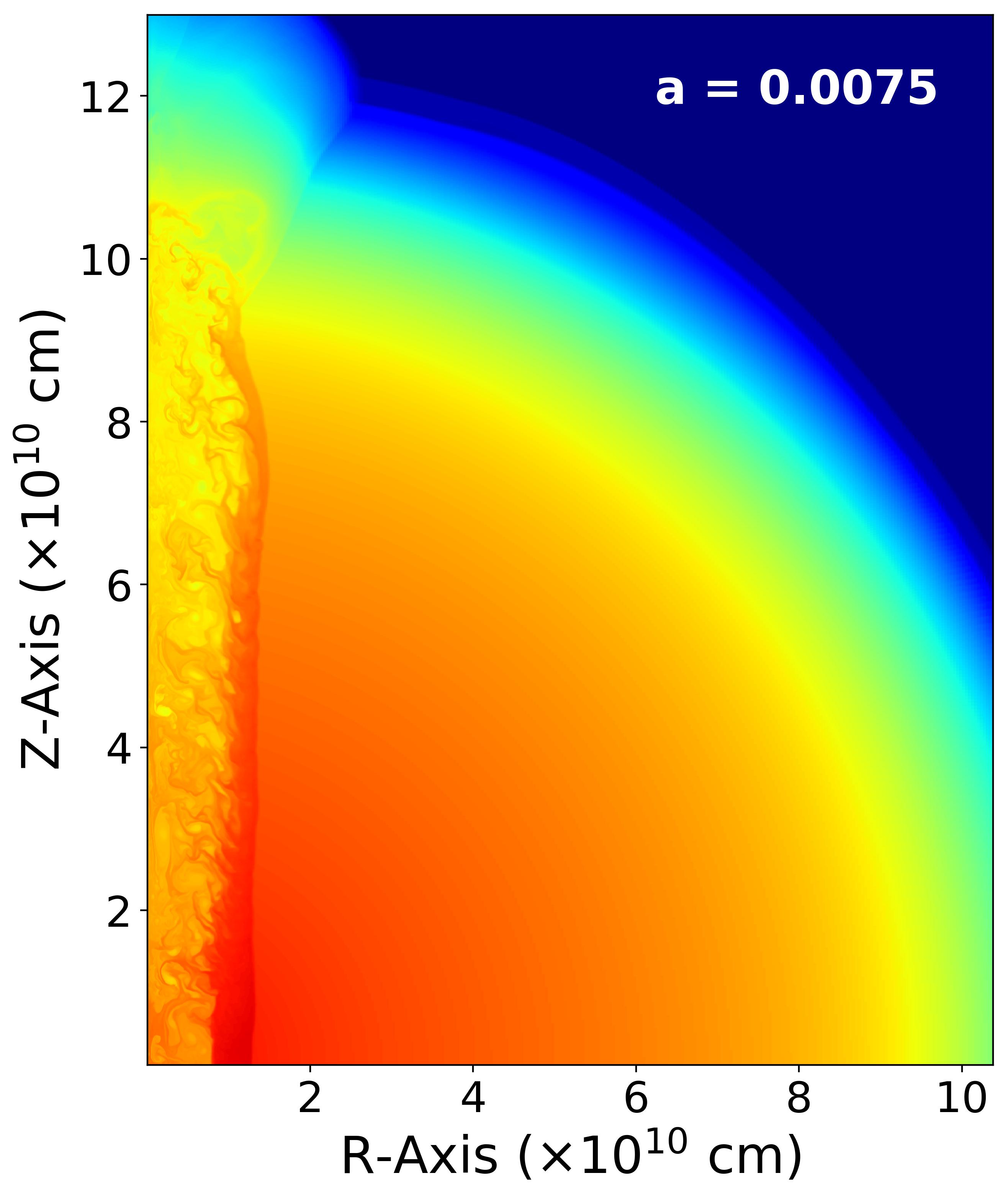}{0.308\textwidth}{(a)}
\fig{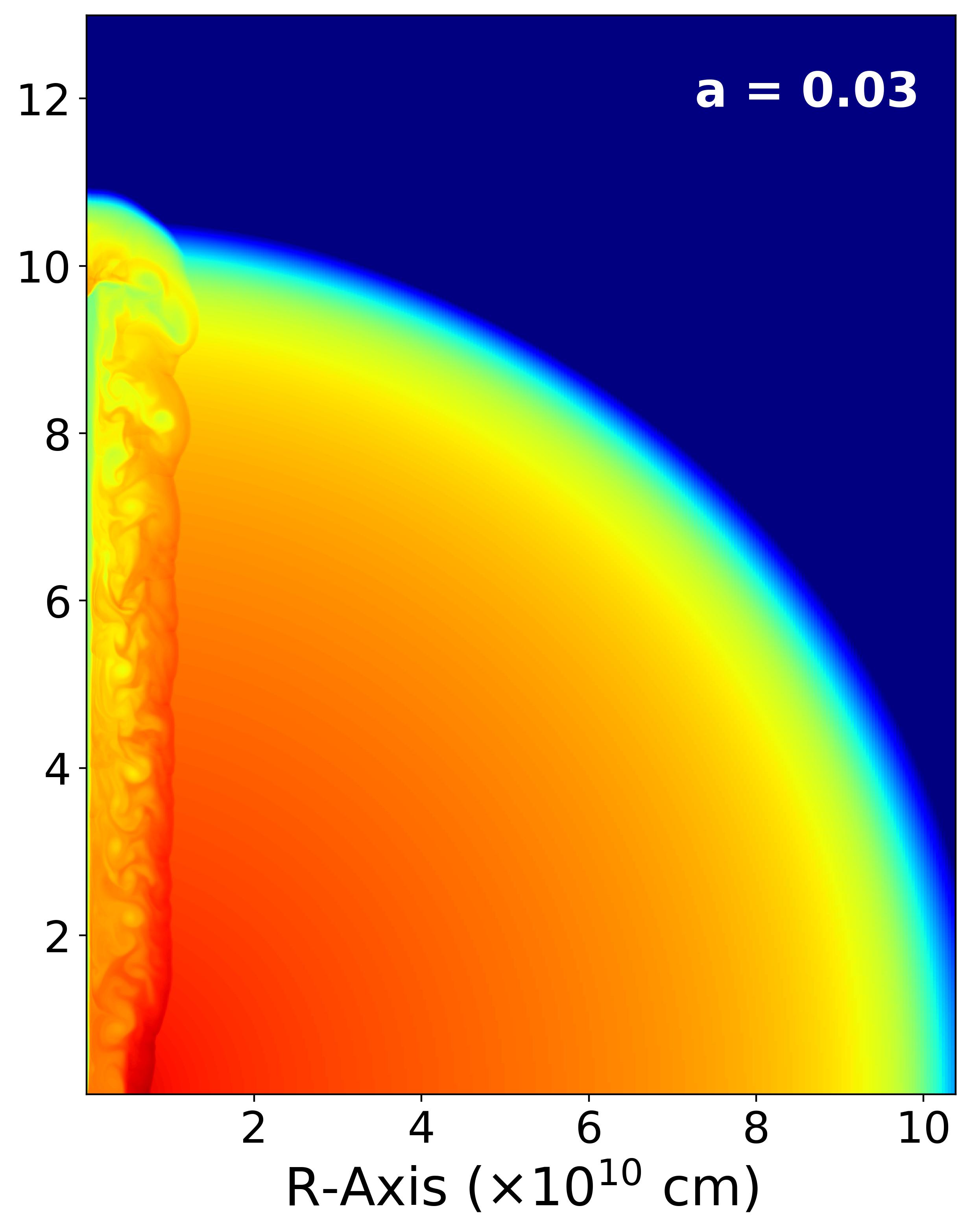}{0.29\textwidth}{(b)}
\fig{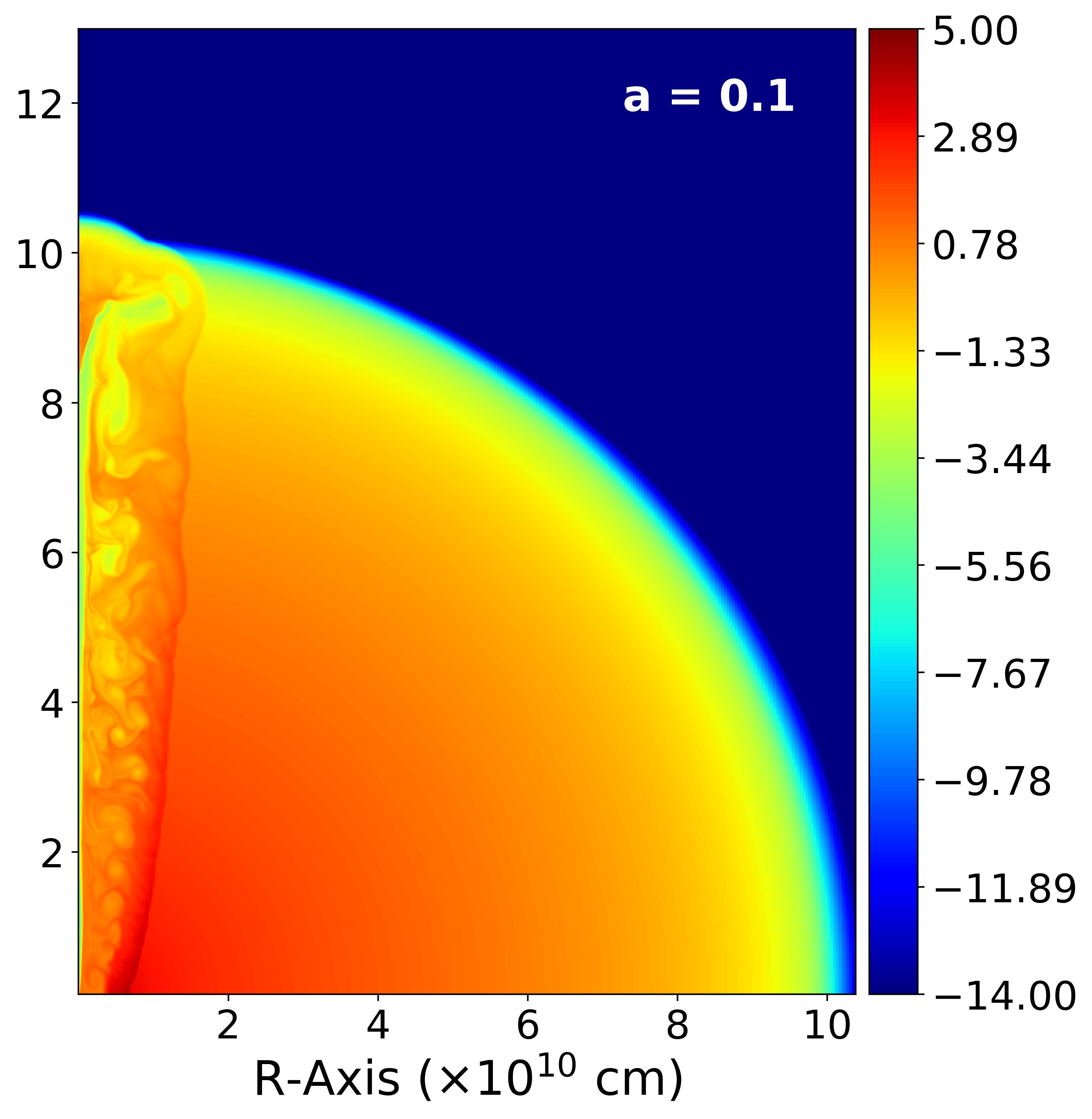}{0.36\textwidth}{(c)}
}

\vspace{0.4cm}

% -------- Bottom row --------
\gridline{
\fig{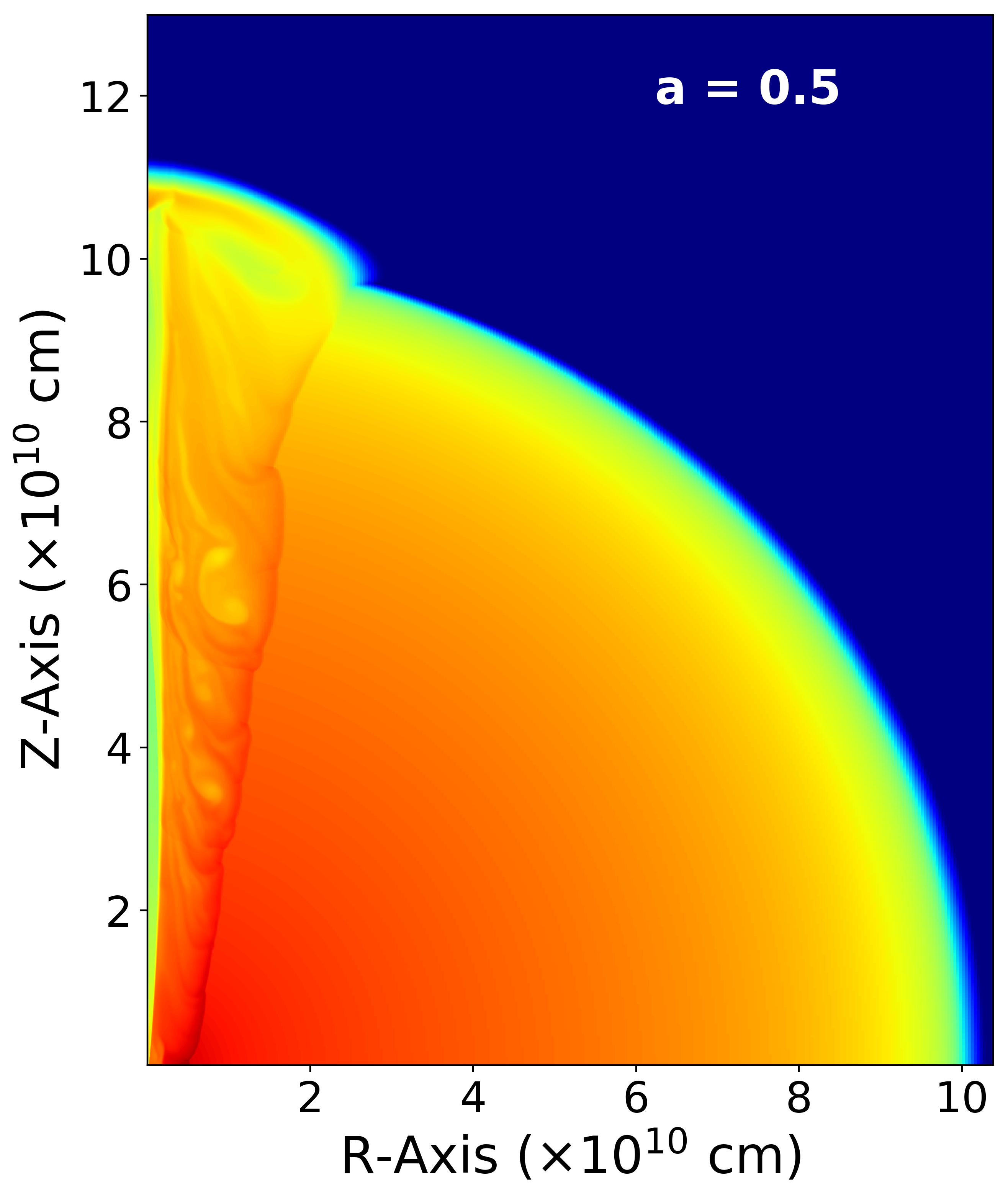}{0.308\textwidth}{(d)}
\fig{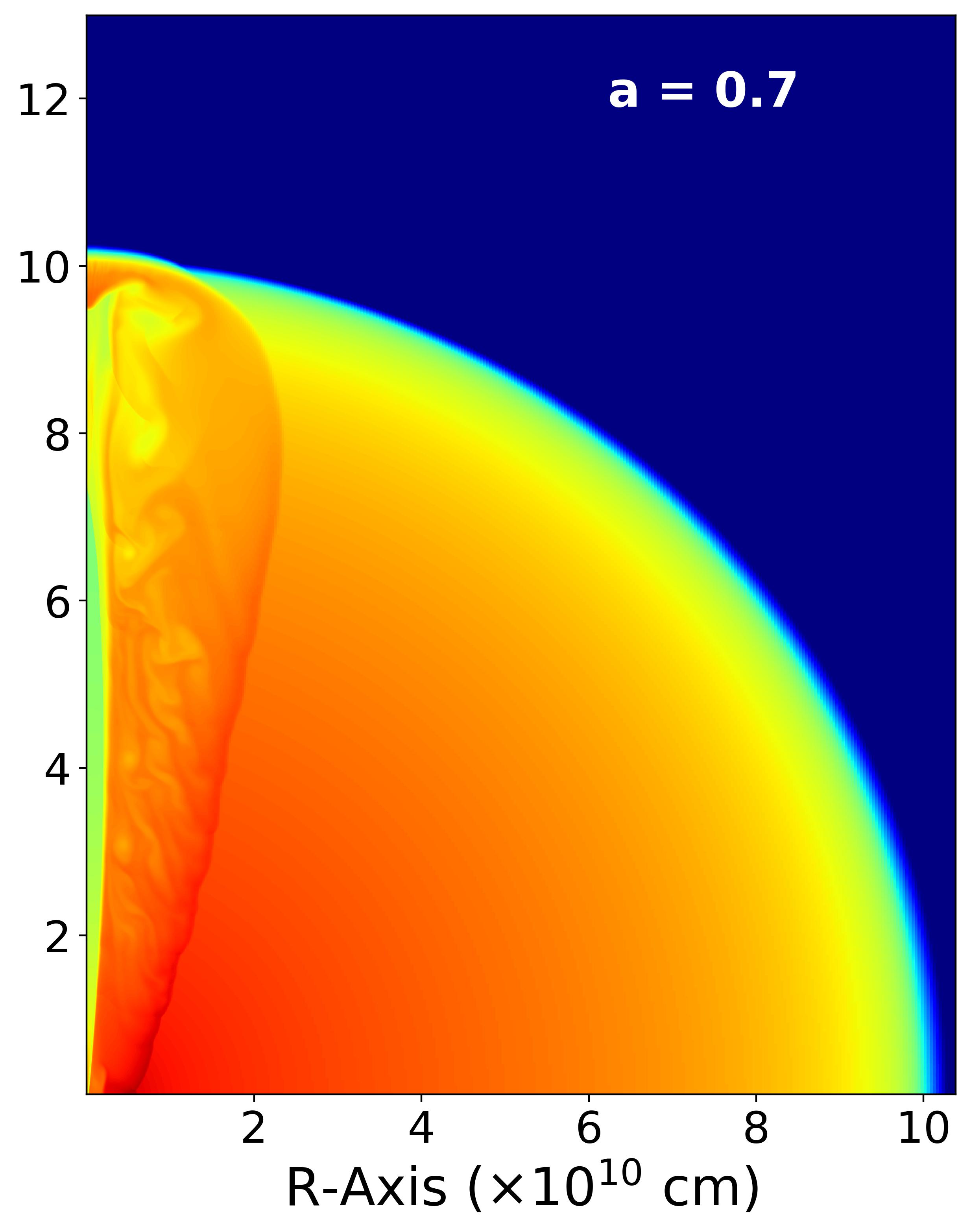}{0.29\textwidth}{(e)}
\fig{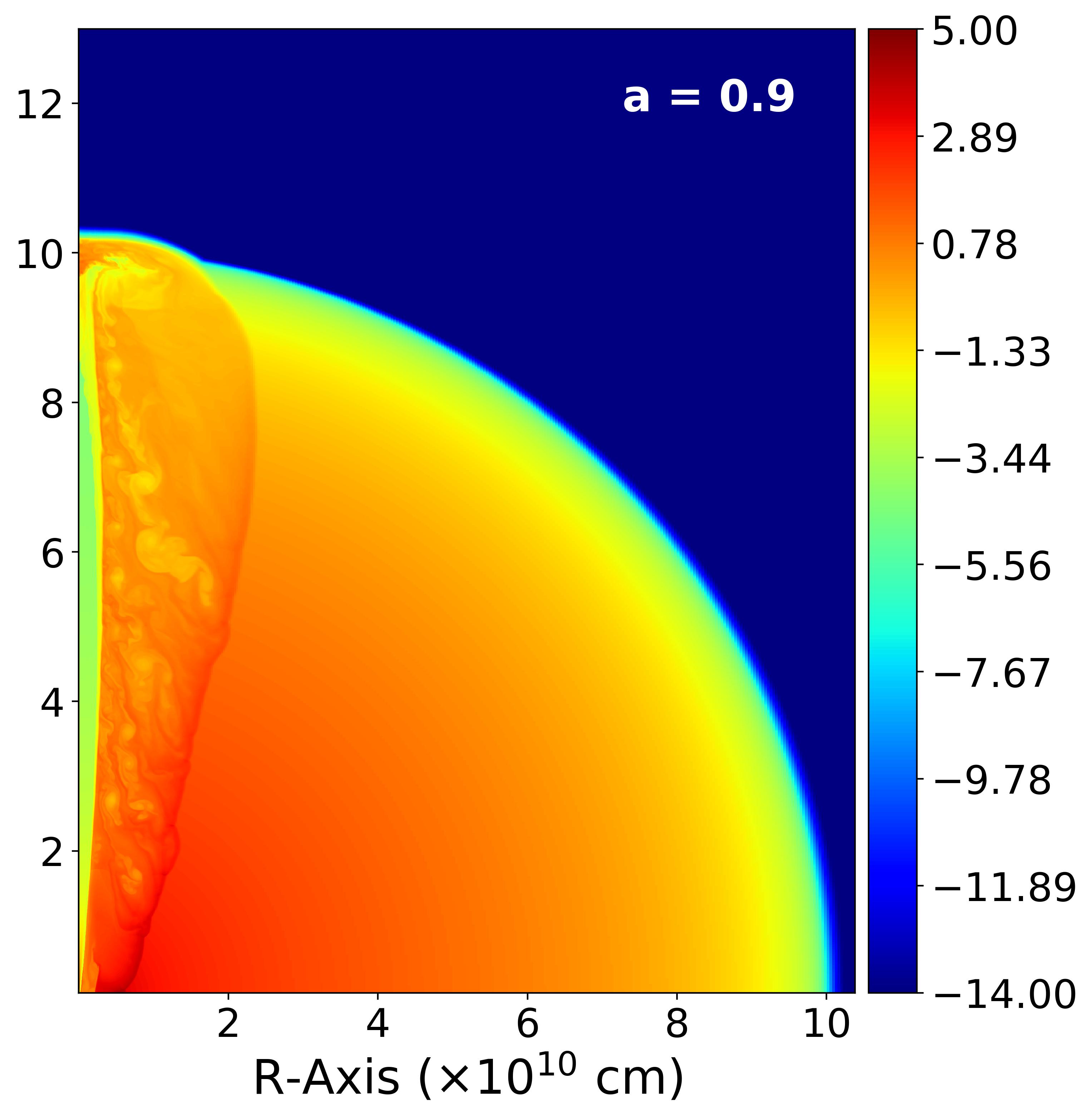}{0.36\textwidth}{(f)}
}

\caption{
Logarithmic density maps (in $\mathrm{g\,cm}^{-3}$) of collapsar jets at breakout time from a $25\,M_{\odot}$ Wolf--Rayet progenitor star for different black hole spins. 
\textbf{Top row (a–c):} Newtonian regime with spins $a = 0.0075$, $0.03$, and $0.1$. 
\textbf{Bottom row (d–f):} Relativistic regime with spins $a = 0.5$, $0.7$, and $a = 0.9$. 
The colorbar corresponds to $\log_{10}(\rho)$.
}

\label{fig:rho_spins_combined_25}

\end{figure*}

We present here the corresponding results for the more massive $25\,M_{\odot}$ progenitor, for which the stellar free-fall timescale is $t_{\rm ff} \sim 550\,\mathrm{s}$. Figure~\ref{fig:rho_spins_combined_25} shows the logarithmic density maps of the jet evolution at breakout (or at $t_{\rm ff}$ for failed cases) across the same range of black hole 
spins considered in the main text. The overall jet morphology and its dependence on spin are qualitatively similar to the $10\,M_{\odot}$ case, with higher-spin systems producing more powerful, well-collimated jets that efficiently carve a low-density polar channel. However, due to the larger stellar envelope, the jet propagation is generally slower, resulting in systematically longer breakout times.

The variation of the normalized breakout time, $t_{\rm B}/t_{\rm ff}$, as a function of spin is shown in Figure~\ref{fig:Chocked_Jet_spin}(b), and the corresponding values are listed in Table~\ref{tab:combined}. As in the 
lower-mass progenitor, we observe a transition from successful breakout to jet choking across the explored spin range. Jets with $a \gtrsim 0.01$ successfully emerge from the star, although with larger $t_{\rm B}/t_{\rm ff}$ compared to the $10\,M_{\odot}$ case. As the spin decreases, the breakout time increases steadily, and for spins of a few 
$\times 10^{-3}$, it exceeds $\sim 20\%$ of the stellar free-fall timescale, indicating significantly less efficient jet propagation. In these low-spin cases, the jet head remains 
deeply embedded within the stellar envelope for an extended period, with much of the injected energy contributing to the expansion of the surrounding stellar material rather than driving rapid outward penetration. These results indicate that more massive progenitors require relatively higher jet power (or equivalently higher black hole spin) to achieve successful breakout and are more susceptible to producing choked jets at low spins.

% \subsection{Calibration of the Analytical Model}
% \label{sec:appendix_25Msun_calibration}

\subsection{Calibration of the Analytical Model}
\label{sec:appendix_25Msun_calibration}

% \begin{figure}
% \centering
% \fig{breakout_calibrated_25.pdf}{0.55\columnwidth}{}
% \caption{
% Jet breakout time $t_B$ as a function of luminosity $L_j$ for a
% $25\,M_\odot$ progenitor. Red circles denote results from numerical
% simulations. The solid dark green line indicates the calibrated model
% using the mean calculated calibration constant, while the blue dashed
% line represents the standard analytical case ($N_s = 1$). The orange
% dotted and purple dash-dotted curves correspond to the minimum and
% maximum $N_s$ values, respectively. The horizontal dotted line
% represents the light-travel limit $R/c$.
% }
% \label{fig:breakout_calibrated_25}
% \end{figure}

\begin{figure*}[t]
\centering

\gridline{
\fig{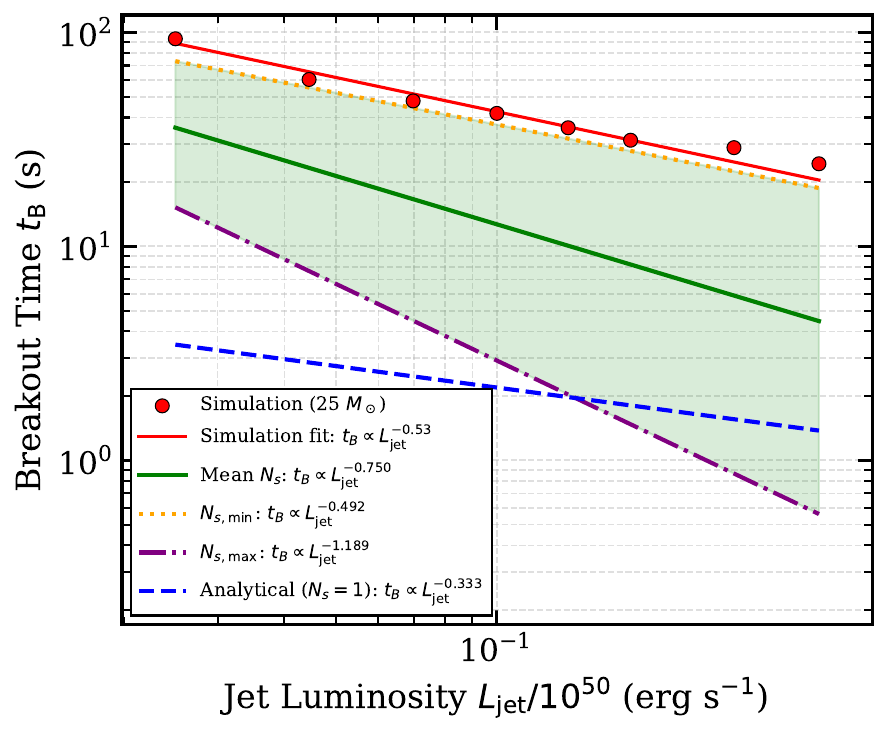}{0.45\textwidth}{(a)}
\fig{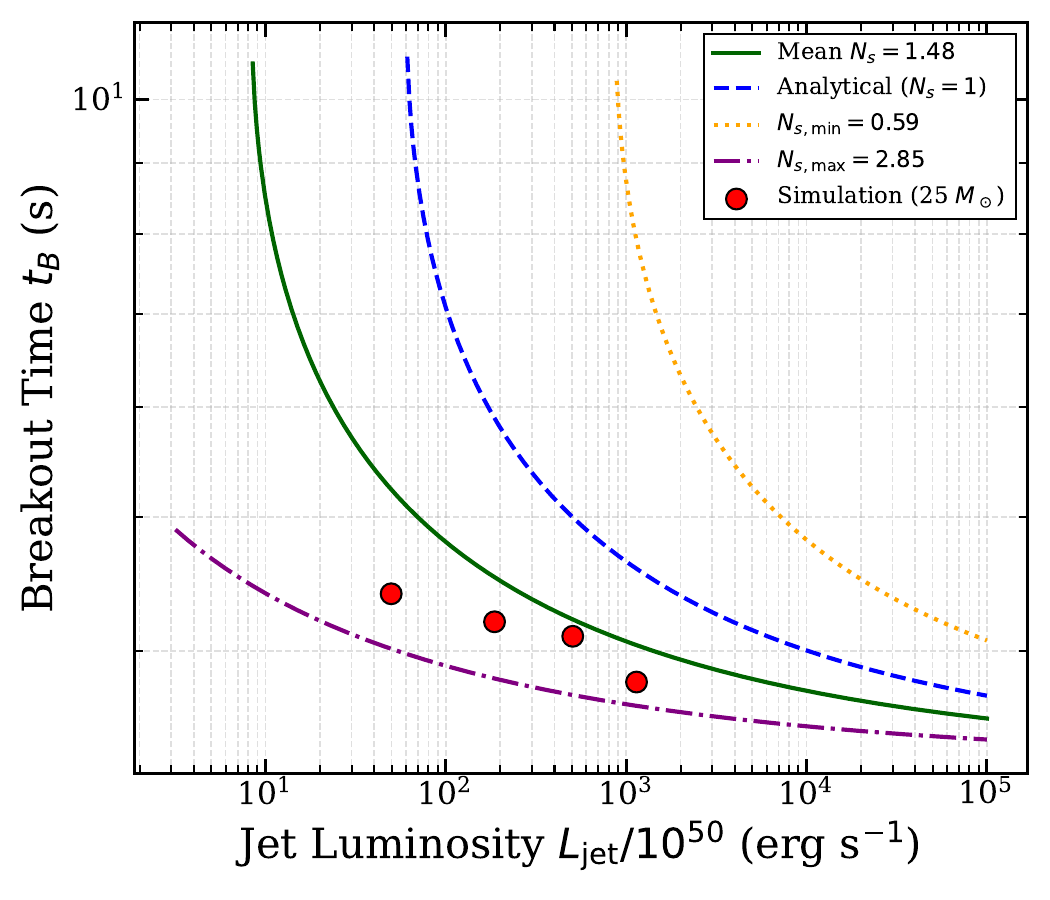}{0.45\textwidth}{(b)}
}
\caption{
% \textbf{(a)} Jet breakout time $t_B$ as a function of jet luminosity in the low-luminosity Newtonian regime. Blue squares denote the numerical simulation results, while the solid blue line represents the power-law fit to the simulations. The red dotted line shows the breakout-time relation obtained from the calibration analysis, while the shaded region represents the corresponding bounds arising from the uncertainties in the calibration.
% \textbf{(b)} Jet breakout time $t_B$ as a function of jet luminosity
% $L_j$ for the $10\,M_\odot$ progenitor. Red circles denote the numerical
% simulation results. The solid dark green line indicates the calibrated
% model using the mean calculated calibration constant, while the blue
% dashed line represents the standard analytical case ($N_s=1$). The
% orange dotted and purple dash-dotted curves correspond to the minimum
% and maximum $N_s$ values, respectively. The horizontal dotted line
% represents the light-travel limit $R/c$.
\textbf{(a)} Jet breakout time $t_B$ versus jet luminosity $L_{jet}$ for the $25\,M_\odot$ progenitor in the low-luminosity Newtonian regime. Red circles show the simulations and the solid red line shows their power-law fit. The solid dark green, blue dashed, orange dotted, and purple dash-dotted lines show the power-law relations from the calibration analysis for the mean, $N_s=1$, minimum, and maximum $N_s$, respectively, with the shaded region indicating the calibration uncertainty. \textbf{(b)} Same as (a), but for the relativistic regime, where the corresponding lines show the model curves.}
\label{fig:breakout_calibrated_25}
\end{figure*}

We find that incorporating the calibration factor into the Newtonian regime yields $\beta_h \propto \tilde{L}_a^{0.75\pm0.13}$ and, consequently, $t_B \propto L_{jet}^{-0.75^{+0.44}_{-0.26}}$. Thus, even after calibration, the predicted luminosity dependence remains systematically steeper than the simulation-derived relation, as discussed in Section~\ref{Newtonian_calibration}. Furthermore, the calibrated analytical scaling does not reproduce the absolute breakout times measured in the simulations (Figure~\ref{fig:breakout_calibrated_25}a). Thus, even after calibrating both the normalization and the power-law index, the analytical model fails to simultaneously capture the $L_{jet}$-dependence and the absolute values of the breakout times obtained from the simulations.

% We find that incorporating the calibration factor into the Newtonian regime yields $\beta_h \propto \tilde{L}_a^{0.75\pm0.13}$ and consequently $t_B \propto L_j^{-0.75^{+0.44}_{-0.26}}$. Thus, even after calibration, the predicted luminosity dependence remains systematically steeper than the simulation-derived relation, as discussed in Section~\ref{Newtonian_calibration}.

In the relativistic regime, the analytical breakout-time relation derived in
Section~\ref{calibration_relativistic} is given by
\begin{equation}
t_B = \frac{R}{c} + k L_{jet}^{-1/5}t_B^{7/5},
\end{equation}
where $k$ depends on the ambient density and jet properties. The numerical
simulations, however, systematically deviate from the standard analytical
prediction obtained by adopting the uncalibrated relation for
$\tilde{L}$. We therefore account for this discrepancy using the calibration
factor $N_s$, defined through the ratio between the simulated and analytical
values of $\tilde{L}$ (Equation~\ref{eq:Ns}).

Including this calibration factor modifies the breakout-time relation to
\begin{equation}
t_B = \frac{R}{c}
+ \frac{k}{N_s}L_{jet}^{-1/5}t_B^{7/5}.
\label{eq:tB_calibrated_25}
\end{equation}
For the $25\,M_\odot$ progenitor, we determine $N_s$ from the simulations in
the relativistic regime ($a>0.2$). The resulting values show a systematic
dependence on the jet luminosity, and are therefore not strictly constant
across the explored luminosity range. The mean value is
$N_s \simeq 1.49$, which we adopt as the fiducial calibration factor for the
analytical model.

Figure~\ref{fig:breakout_calibrated_25}b compares the numerical breakout
times with the corresponding analytical predictions. The standard
analytical model with $N_s=1$ systematically overestimates the breakout
time. In contrast, incorporating the mean calibration factor produces a
substantially better agreement with the numerical results. To illustrate
the effect of the variation in $N_s$, we additionally show the predictions
obtained using the minimum and maximum values of $N_s$ measured from the
simulations. These curves approximately bracket the range of breakout times
predicted by the calibrated model.

The comparison demonstrates that the discrepancy between the standard
analytical prediction and the numerical results in the relativistic regime
can be largely accounted for by the calibration of the dimensionless
$\tilde{L_s}$ through $N_s$. The calibrated relation therefore provides a more
accurate description of the breakout-time dependence for the
$25\,M_\odot$ progenitor.

% \begin{figure*}
% \centering

% % -------- Top row --------
% \begin{subfigure}[t]{0.308\textwidth}
%     \includegraphics[width=\linewidth]{rho_spin_0.0075_25.jpg}
% \end{subfigure}
% \hfill
% \begin{subfigure}[t]{0.29\textwidth}
%     \includegraphics[width=\linewidth]{rho_spin_0.03_25.jpg}
% \end{subfigure}
% \hfill
% \begin{subfigure}[t]{0.36\textwidth}
%     \includegraphics[width=\linewidth]{rho_spin_0.1_25.jpg}
% \end{subfigure}

% \vspace{0.4cm}

% % -------- Bottom row --------
% \begin{subfigure}[t]{0.308\textwidth}
%     \includegraphics[width=\linewidth]{rho_spin_0.5_25.jpg}
% \end{subfigure}
% \hfill
% \begin{subfigure}[t]{0.29\textwidth}
%     \includegraphics[width=\linewidth]{rho_spin_0.7_25.jpg}
% \end{subfigure}
% \hfill
% \begin{subfigure}[t]{0.36\textwidth}
%     \includegraphics[width=\linewidth]{rho_spin_0.9_25.jpg}
% \end{subfigure}

% \caption{Logarithmic density maps (in $\mathrm{g\,cm}^{-3}$) of collapsar jets at breakout time from a $25\,M_{\odot}$ Wolf--Rayet progenitor star for different black hole spins. 
% \textbf{Top row:} Newtonian regime with spins $a = 0.0075$, $0.03$, and $0.1$. 
% \textbf{Bottom row:} Relativistic regime with spins $a = 0.5$, $0.7$, and $0.9$. 
% The colorbar corresponds to $\log_{10}(\rho)$. }
% \label{fig:rho_spins_combined_25}

% \end{figure*}

\section{Generic Relation between breakout time and injection jet luminosity}
\label{generic_relation}

To obtain a generic relation between the jet breakout time, $t_B$, and the injection luminosity $L_{jet}$, we generalize the jet-head velocity ($\beta_h$) scaling by allowing its dependence on the dimensionless parameter $\tilde{L}_s$ to be characterized by 
an arbitrary exponent $\delta$
\begin{equation*}
\beta_h \approx \tilde{L}_s^{\delta},
\end{equation*}
using $\tilde{L}_s=\tilde{L}_a\, N_s^2$ and parameterizing the calibration factor as a power-law function of $\tilde{L}_a$, 
\begin{equation*}
    N_s=A\, \tilde{L}_a^p \, ,
\end{equation*}
the jet-head velocity can be expressed 
in terms of $\tilde{L}_a$, $A$, $p$, and $\delta$, 
\begin{equation*}
    \beta_h \approx A^{2\delta} \, \tilde{L}_a^{\delta (2p+1)}.
\end{equation*}
Substituting this generalized velocity scaling into the analytical expression for $\tilde{L}_a$ (Equation~\ref{eq:collimated}) and carrying out the standard integration of the jet-head propagation time to the stellar 
surface, $R_s$, yields a general expression for the breakout time as a function of the jet injection luminosity. The resulting expression, including its dependence on the progenitor properties and the parameters $A$, $p$, and $\delta$, is given by

\begin{equation}
    t_B \approx \left[\frac{R_s \, \left(1 - \frac{4 \delta(2p+1}{5}\right)}{c \, A^{2\delta} \, {\cal{D}}^{\delta(2p+1)}}\right]^{\frac{1}{1-4\delta(2p+1)/5}} \, L_{jet}^{\frac{-2\delta(2p+1)/5}{1-4\delta(2p+1)/5}},
\end{equation}
where 
\begin{equation*}
    {\cal{D}} = \left(\frac{16 \Omega}{3\pi \theta_0^4 \rho_a \, c^5}\right)^{2/5} ,
\end{equation*}
with $\rho_a$ representing the ambient density and $c$ denote the speed of light. 

\bibliography{ref.bib}{}
\bibliographystyle{aasjournal}



\end{document}